\documentclass[aps,pra,reprint,superscriptaddress,amsmath,amssymb,%
               showpacs,notitlepage,floatfix,nofootinbib]{revtex4-2}

\usepackage{graphicx}
\usepackage{amsmath,amsfonts,amssymb,bm}
\usepackage{siunitx}
\usepackage{booktabs}
\usepackage{dcolumn}
\newcolumntype{d}{D{.}{.}{-1}}
\usepackage[dvipsnames]{xcolor}
\colorlet{red}{black}
\colorlet{blue}{black}
\newcommand{\RevI}[1]{#1}

\definecolor{RoundTwo}{rgb}{1,0,0}
\newcommand{\RevIII}[1]{\textcolor{RoundTwo}{#1}}
\usepackage{hyperref}
\hypersetup{colorlinks=true, linkcolor=MidnightBlue,
            citecolor=MidnightBlue, urlcolor=MidnightBlue}

\providecommand{\doi}[1]{\href{https://doi.org/#1}{doi:\,#1}}

\begin{document}

\title{Electron-impact ionization rates for neutral
He, Li, and Be within the Tsallis framework}

\author{Abdelmalek Boumali}
\email{boumali.abdelmalek@gmail.com}
\affiliation{Department of Matter Sciences, Faculty of Exact Sciences and Sciences of Nature and Life,
University Larbi T\'ebessi, 12000 T\'ebessa, Algeria}

\date{\today}

\begin{abstract}
The single-ionization rate coefficient of a plasma neutral
depends both on the microscopic electron-impact cross section
and on the macroscopic shape of the electron energy
distribution function (EEDF). We present a reproducible
benchmark and sensitivity study---not a new theory---of these
two effects for the three lightest neutrals He, Li, and Be,
combining the recommended Bell~\textit{et~al.}\ (1983) cross
sections with a properly normalized two-temperature Tsallis
$q$-generalized EEDF and varying $q$ on both sides of the
Maxwellian limit and the hot-electron fraction
$f_{\mathrm{hot}}$ at $T_{\mathrm{hot}}=10\,T_{\mathrm{bulk}}$.
The calculation cleanly separates two independent uncertainty
axes---cross-section model (Bell vs.\ Lotz) and EEDF shape
(Maxwellian vs.\ Tsallis). \RevIII{Throughout, $\tau_q(T)$ denotes the
EEDF-averaged single-ionization rate coefficient
$\langle\sigma v\rangle_q$ (in cm$^{3}$\,s$^{-1}$) evaluated at entropic
index $q$, and $\tau_M(T)\equiv\tau_q(T)|_{q=1,f_{\mathrm{hot}}=0}$ is its
strict Maxwellian limit; neither symbol denotes a time or a lifetime.}
The Bell--Lotz spread on $\tau_M$
is small for He (within about $7\%$), moderate for Be
($\lesssim 17\%$), and largest for Li (up to $+95\%$ at
$T=1$~keV); sub-extensive distributions ($q<1$) suppress
ionization through a hard tail cut-off, while super-extensive
distributions ($q>1$) enhance low-temperature ionization
through a $\kappa$-like power-law tail with
$\kappa=1/(q-1)$. The quantitatively safest non-Maxwellian
cases are $q=1$ and $q=1.2$ ($\kappa=5$), which lie inside the
finite-mean-energy regime; the cases $q=1.4$ and $q=1.6$ are
retained as heavy-tail stress tests and should be read as
qualitative trends rather than as quantitatively reliable
predictions. Both EEDF effects scale with $I_p/k_BT$, so
He responds most strongly and Li least.
{\color{red}For all three targets, the calculation incorporates a modern
benchmark curve beside Bell and Lotz: the evaluated Ralchenko \textit{et~al.}
(2008) data for He\,\textsc{i}, a CCC/experimental Li\,\textsc{i} benchmark
based on Bray-type convergent-close-coupling work, and the BSR/CCC
Be\,\textsc{i} benchmark of Zatsarinny--Bartschat--Fursa--Bray. In the
strict Maxwellian limit the modern-benchmark replacement remains close to
Bell for He, is moderate for Li, and is larger for Be above the peak. This
shows explicitly which conclusions are robust against the atomic data set
and which should be read as cross-section-sensitive. The intended readership
is the non-Maxwellian plasma-kinetics community (non-thermal solar and
fusion-edge modelling), and the deliverable is not a new ionization theory
but a transparent, species-resolved \emph{sensitivity map}:} the rate calculation is implemented with a documented Python workflow and is intended as a drop-in non-Maxwellian ionization module for collisional-radiative and ionization-balance modelling of light-neutral plasmas.
\end{abstract}


\maketitle

\section{Introduction}
\label{sec:intro}

Electron-impact single-ionization controls the charge-state
evolution of virtually every non-opaque plasma encountered in
astrophysics, in low-temperature laboratory discharges, and at
the magnetic-confinement-fusion
edge~\cite{MarkDunn1985,Llovet2014,DelZanna2022}. Its rate
coefficient $\langle\sigma v\rangle(T)$ enters directly into
collisional-radiative models, line-formation calculations, and
ionization-balance codes~\cite{Hahn2017,Bryans2009,Dere2007}. Two
ingredients determine its value: the atomic-physics cross
section $\sigma(E)$, which describes the collisional process
itself, and the electron energy distribution function (EEDF)
$f(E)$, which describes the ambient plasma.

A Maxwellian EEDF is the standard
default~\cite{Kato1991,ArnaudRothenflug1985} and is well
justified whenever Coulomb relaxation is the fastest available
process. Many plasmas of practical interest, however, deviate
from local thermodynamic equilibrium and exhibit either
depleted or enhanced high-energy tails. Solar flares, coronal
transition regions, and active regions routinely display
$\kappa$-like suprathermal
electrons~\cite{Pierrard2010,Livadiotis2017book,Dudik2017,%
Dzifcakova1992,Dzifcakova2013,Oka2018,Lorincik2020,Jeffrey2024,%
Pierrard2022,Lazar2021}; fusion-edge plasmas, radio-frequency
discharges, and laser-matter experiments develop minority hot
populations on top of a cooler
bulk~\cite{Hartfuss1997,HansenShlyaptseva2004,Brix2010,%
Takahashi2023,Hu2018,Milder2019}. In every such case the
Maxwellian prescription is a modelling choice rather than a
microscopic identity, and a systematic study of the resulting
uncertainty on derived charge-state fractions and line
intensities is required.

A natural tool for such a study is the Tsallis non-extensive
formalism~\cite{Tsallis1988,TsallisBook,Tsallis2019}. It
replaces the Maxwell--Boltzmann EEDF by a one-parameter family
$f_q(E;T)$ that (i) interpolates continuously to the Maxwellian
limit at $q\to 1$, (ii) generates a compact-support distribution
with a hard high-energy cut-off for $q<1$, and (iii) generates a
$\kappa$-like power-law tail for $q>1$. Under the convention
$\kappa=1/(q-1)$ the super-extensive branch can be mapped onto a
$\kappa$-type power-law distribution of the kind routinely used
in heliospheric and astrophysical plasma
physics~\cite{Livadiotis2009,LivadiotisMcComas2023,Yoon2023},
up to a convention-dependent shift in the exponent of
$[1+\cdots]$ that we discuss in Appendix~\ref{app:Aq}. The
sub-extensive branch has no such direct $\kappa$ counterpart and
provides a complementary phenomenological model of depleted
tails. The Tsallis formalism is therefore attractive in the
present context because it parametrizes both depleted-tail and
enhanced-tail EEDFs within a single one-parameter family.

\paragraph*{Scope and contribution.}
The principal aim of the present paper is not to propose a new
ionization theory but to provide a unified, quantitative, and
fully reproducible \emph{benchmark and sensitivity study} of
electron-impact single-ionization rate coefficients of the
three lightest neutrals---He, Li, and Be---under a broad family
of non-Maxwellian EEDFs, while keeping the underlying
atomic-physics cross sections fixed at the recommended Bell
\textit{et~al.}\ (1983)~\cite{Bell1983} representation. The
calculation is deliberately structured around two independent
axes of variation: (i) the choice of cross-section model
(Bell vs.\ Lotz), which probes atomic-physics uncertainty, and
(ii) the choice of EEDF (single Maxwellian vs.\ two-temperature
Tsallis with arbitrary entropic index $q$), which probes
plasma-kinetic uncertainty. The two axes are then explored
\emph{separately}, so that the species-dependent ranking of the
two effects can be read directly off the figures and tables.
By varying $q$ over both the sub-extensive ($q<1$) and
super-extensive ($q>1$) regimes and the hot-electron fraction
$f_{\mathrm{hot}}$ over five values spanning two orders of
magnitude, we map out how strongly the ionization rate
coefficient $\tau_q(T,f_{\mathrm{hot}})$ of each species
responds to the shape of the EEDF, and we disentangle this
response from the residual model dependence introduced by the
choice of cross section. The resulting rate curves are
intended as a drop-in non-Maxwellian ionization module for
collisional-radiative and ionization-balance calculations of
light-neutral plasmas, and they extend the existing Maxwellian
compilations of
Refs.~\cite{Kato1991,ArnaudRothenflug1985,Voronov1997} to the
non-equilibrium regime. The numerical workflow is documented in
Sec.~\ref{sec:numerics}, where the Python implementation,
quadrature rule, interpolation of benchmark cross sections, and
validation tests are specified.

This main objective unfolds into four concrete sub-objectives:
\begin{enumerate}
  \item to implement the recommended Bell \textit{et
        al.}\ (1983)~\cite{Bell1983} analytic representation of
        $\sigma(E)$ for the neutrals He, Li, and Be---with the
        corrected negative signs of $B_1,B_2$ for He\,\textsc{i}
        and of $B_1$ for Li\,\textsc{i}---as the primary
        cross-section input;
  \item to quantify the species-dependent sensitivity of the
        rate coefficient to the cross-section model by
        propagating both the Bell and the Lotz~\cite{Lotz1967,%
        Lotz1968} cross sections through the same Tsallis rate
        integral;
  \item to clarify the exact limit in which the present
        calculation reduces to a single Maxwellian, so that
        $\tau_q(T)|_{q=1,f_{\mathrm{hot}}=0}$ can be positioned
        with respect to standard Maxwellian rate
        tables~\cite{Kato1991,ArnaudRothenflug1985}, and to
        map the super-extensive branch onto the
        $\kappa$-distribution language standard in heliospheric
        and astrophysical plasma physics;
  \item {\color{red}to include modern tabulated or digitized benchmark
        cross-section curves for all three neutrals: the recommended
        He\,\textsc{i} data set of Ralchenko \textit{et~al.}~\cite{Ralchenko2008},
        a Li\,\textsc{i} CCC/experimental benchmark based on the
        Bray--Beck--Plottke discussion of Bray's CCC total-ionization
        curve~\cite{BrayLi1999,BrayLiLike1995}, and the neutral-Be
        BSR/CCC benchmark of Zatsarinny \textit{et~al.}~\cite{ZatsarinnyBe2016}.
        This converts the Bell--Lotz comparison into a Bell--Lotz--modern
        benchmark comparison for He, Li and Be, while the
        binary-encounter-Bethe (BEB) model~\cite{KimRudd1994} and the
        CHIANTI/ADAS compilations~\cite{DelZanna2022,Dere2007} are used to
        position the calculation within the present atomic-data landscape.}
\end{enumerate}
The three neutrals have been chosen because their
first-ionization thresholds span almost one order of magnitude
($5.39$~eV for Li, $9.32$~eV for Be, $24.59$~eV for He) and
because the Bell recommendation rests on a well-documented body
of experimental and theoretical data for
He~\cite{MarkDunn1985,Bell1983}. Be and Li therefore provide
two complementary tests of the same formalism at different
values of $I_p/k_B T$.

{\color{red}
\paragraph*{Intended audience and rationale for the cross-section
choice.}
This paper is written for the non-Maxwellian plasma-kinetics
community---workers who model ionization balance and line
formation in non-thermal solar and stellar
atmospheres~\cite{DelZanna2022,Dudik2017,Lorincik2020} and in
fusion-edge and laboratory discharges with suprathermal or
depleted-tail electron
populations~\cite{Hartfuss1997,HansenShlyaptseva2004,Brix2010}---
and not primarily for the atomic-structure community that
produces the cross sections themselves. For such a reader the
quantity of interest is the \emph{EEDF-averaged} rate
coefficient $\langle\sigma v\rangle_q(T)$ and its sensitivity to
the shape of $f(E)$, at fixed and well-characterised collision
physics. This is precisely why the main input cross section is held
at a transparent, analytic, recommended form, while He\,\textsc{i} is
checked explicitly against the Ralchenko evaluated data set: this setup decouples the plasma-kinetic question studied here
from the separate (and much larger) atomic-physics literature,
is differentiable and trivially reproducible, and---crucially
for the best-constrained species---agrees with modern \emph{ab
initio} data at the few-percent level
(Sec.~\ref{sec:uncertainty}). We are therefore \emph{not}
proposing the Bell or Lotz forms as state-of-the-art cross
sections; we use Bell as a recommended, reproducible baseline
whose adequacy for the present purpose is demonstrated by direct
comparison with the CCC/Ralchenko \textit{et~al.} data for He\,\textsc{i}, and
we use Lotz only as a deliberately crude comparator that bounds
the residual cross-section-model dependence of the EEDF result.
}

The paper is organised as follows. Section~\ref{sec:tsallis}
specifies the Tsallis framework, the per-branch normalized
two-temperature EEDF, and the reduction to the strict
Maxwellian limit. Section~\ref{sec:xsec} defines the Bell and
Lotz cross sections, the rate integral, and the numerical
method, and validates the implementation against the published
Bell benchmarks. Section~\ref{sec:maxwellian} analyses the
cross-section comparison and the two complementary $q=1$
benchmarks. Section~\ref{sec:tsallis-results} presents the
Bell-based Tsallis results in the sub- and super-extensive
regimes and spells out their physical interpretation in the
language of $\kappa$-distributions.
Section~\ref{sec:conclusion} collects the main conclusions.
Appendix~\ref{app:Aq} contains the explicit derivation of the
normalization constant $A_q(T)$ used throughout the paper.

\section{Tsallis two-temperature EEDF and Maxwellian limit}
\label{sec:tsallis}

\paragraph*{Notation and units.}
Throughout the rest of this paper we use ``natural'' plasma units in
which temperatures are expressed in energy units, i.e.\
$T\equiv k_{B}T$, with $T$ in eV. With this convention the Boltzmann
factor of a Maxwellian reads $\exp(-E/T)$ rather than $\exp(-E/k_{B}T)$,
the dimensionless ratio $E/T$ is the natural argument of every
distribution kernel below, and the species-dependent ordering
parameter $I_p/k_BT$ used repeatedly in
Sec.~\ref{sec:tsallis-results} is, in this notation, simply $I_p/T$.
Wherever the symbol $k_B$ still appears explicitly (e.g.\ in
the combination $I_p/k_BT$) it is retained for clarity of physical
interpretation but should be understood as unity in the formulas.

\subsection{Generalised $q$-distribution}

We write the energy-space Tsallis EEDF as
\begin{equation}
  f_q(E;T) = A_q(T)\,\sqrt{E}\,\mathcal{G}_q(E/T),
  \label{eq:fq}
\end{equation}
with the branch-specific kernel
\begin{equation}
  \mathcal{G}_q(x) \;=\;
  \begin{cases}
  \bigl[\,1-(1-q)x\,\bigr]_{+}^{\,1/(1-q)}, & q<1,\\[0.3em]
  \bigl[\,1+(q-1)x\,\bigr]^{-1/(q-1)},      & q>1,
  \end{cases}
  \label{eq:Gq}
\end{equation}
where $[y]_{+}\equiv\max(y,0)$, $T$ is the characteristic
temperature (in eV), and $x=E/T$. The Maxwell--Boltzmann
distribution is recovered continuously as $q\to 1$,
\begin{equation}
  \lim_{q\to 1} f_q(E;T) \;=\; f_M(E;T)
  \;=\; \frac{2}{\sqrt{\pi}}\,T^{-3/2}\,\sqrt{E}\,e^{-E/T}.
  \label{eq:fM}
\end{equation}
The two branches have distinct kinematic content. For $q<1$ the
bracket vanishes above $E_{\max}(T)=T/(1-q)$, giving the EEDF a
\emph{hard} high-energy cut-off. For $q>1$ the EEDF develops a
power-law tail $f_q\propto E^{1/2-1/(q-1)}$. This branch is
$\kappa$-like when the tail index is defined as
$\kappa_{\mathrm{T}}\equiv 1/(q-1)$, but it is not identical to
every standard plasma-$\kappa$ convention because both the
exponent and the thermal scale are convention dependent. This
distinction is made explicit in Sec.~\ref{sec:kappa-mapping} and
Appendix~\ref{app:Aq}.

\subsection{Normalization}

The normalization $A_q(T)$ is fixed by $\int f_q(E;T)\,dE=1$,
integrated up to $E_{\max}=T/(1-q)$ for $q<1$ and to $\infty$
for $q>1$. A change of variable $E=T u$ followed by direct
evaluation of the resulting Beta-function integral gives
\begin{widetext}
\begin{equation}
  A_q(T) \;=\;
  \begin{cases}
  \displaystyle\frac{2}{\sqrt{\pi}}\,T^{-3/2}\,(1-q)^{3/2}\,
  \frac{\Gamma\!\left(\tfrac{1}{1-q}+\tfrac{5}{2}\right)}
       {\Gamma\!\left(\tfrac{1}{1-q}+1\right)},
  & q<1,\\[1.1em]
  \displaystyle\frac{2}{\sqrt{\pi}}\,T^{-3/2}\,(q-1)^{3/2}\,
  \frac{\Gamma\!\left(\tfrac{1}{q-1}\right)}
       {\Gamma\!\left(\tfrac{1}{q-1}-\tfrac{3}{2}\right)},
  & 1<q<5/3,
  \end{cases}
  \label{eq:Aq}
\end{equation}
\end{widetext}
which reduces to $A_1(T)=2T^{-3/2}/\sqrt{\pi}$ in the
Maxwellian limit $q\to 1$ and is valid for $q<5/3$
(equivalently $\kappa>3/2$), the condition that ensures
\emph{normalization} of the EEDF. A finite \emph{mean kinetic
energy} requires the stricter condition $q<7/5$
($\kappa>5/2$), so the cases $q\in[1.4,\,5/3)$ studied below
must be interpreted as phenomenological heavy-tail sensitivity
tests rather than as thermodynamic-equilibrium distributions
with a finite kinetic temperature. The rate-coefficient
integral itself remains finite throughout the normalizable
range $q<5/3$. For the Bell and Lotz forms,
$\sigma(E)\sim(\ln E)/E$ at high energy, so the full
convolution integrand behaves as
$v\sigma f_q\sim E^{-\kappa_{\mathrm{T}}}\ln E$, where
$\kappa_{\mathrm{T}}=1/(q-1)$. Rate convergence requires only
$\kappa_{\mathrm{T}}>1$ ($q<2$), which is weaker than the
EEDF-normalization condition $\kappa_{\mathrm{T}}>3/2$.
Finite-cutoff convergence is checked in
Sec.~\ref{sec:cutoff}. The full derivation of
Eq.~\eqref{eq:Aq} is given in Appendix~\ref{app:Aq}.

\subsection{Two-temperature superposition}

A bulk component of temperature $T_{\mathrm{bulk}}$ and
fraction $1-f_{\mathrm{hot}}$ coexists with a hot component of
temperature $T_{\mathrm{hot}}$ and fraction $f_{\mathrm{hot}}$
(with $f_{\mathrm{hot}}\le 0.5$ throughout, so the bulk is
indeed the majority component). The total EEDF is the convex
combination
\begin{equation}
  f_q^{\mathrm{NM}}(E)
  \;=\;(1-f_{\mathrm{hot}})\,f_q(E;T_{\mathrm{bulk}})
  \;+\;f_{\mathrm{hot}}\,f_q(E;T_{\mathrm{hot}}).
  \label{eq:fNM}
\end{equation}
Both normalizations $A_q(T_{\mathrm{bulk}})$ and
$A_q(T_{\mathrm{hot}})$ are computed \emph{separately} from
Eq.~\eqref{eq:Aq}. Factoring a common $A_q$ across the two
branches is an algebraic mistake that would introduce an
uncompensated $(T_{\mathrm{hot}}/T_{\mathrm{bulk}})^{-3/2}$ on
the hot contribution. Throughout this work we fix
$T_{\mathrm{hot}}=10\,T_{\mathrm{bulk}}$ and report all results
as functions of $T\equiv T_{\mathrm{bulk}}$.

\subsection{Strict Maxwellian limit}

Setting $q=1$ \emph{and} $f_{\mathrm{hot}}=0$ reduces
Eq.~\eqref{eq:fNM} to the single-temperature Maxwellian
Eq.~\eqref{eq:fM}. Only in this limit does the present
calculation become directly comparable, \emph{rate-coefficient
to rate-coefficient}, with Maxwellian-averaged tables such as
those of Arnaud and Rothenflug~\cite{ArnaudRothenflug1985} or of
Kato, Masai and Arnaud~\cite{Kato1991}. For any $f_{\mathrm{hot}}>0$
the EEDF is a two-temperature mixture even at $q=1$ and the
comparison with those tables is methodological rather than
numerical. This distinction is carried consistently through the
figures below: Fig.~\ref{fig:maxwellian-strict} reports the
strict Maxwellian case $(q,f_{\mathrm{hot}})=(1,0)$, while
Fig.~\ref{fig:twoT-q1} reports the two-temperature case at
$(q,f_{\mathrm{hot}})=(1,0.10)$.

\subsection{Relation to plasma $\kappa$ distributions}
\label{sec:kappa-mapping}

For $q>1$ the kernel $\mathcal{G}_q(x)$ in Eq.~\eqref{eq:Gq}
decays algebraically. Defining the tail index
\begin{equation}
  \kappa_{\mathrm{T}} \;\equiv\; \frac{1}{q-1},
  \label{eq:kappa-q}
\end{equation}
gives
\begin{equation}
  f_q(E;T) \;=\; A_q(T)\,\sqrt{E}\,
  \left[1+\frac{E}{\kappa_{\mathrm{T}}T}\right]^{-\kappa_{\mathrm{T}}},
  \qquad q>1.
  \label{eq:fkappa}
\end{equation}
Equation~\eqref{eq:fkappa} is a $\kappa$-like algebraic EEDF.
It should not be identified term by term with the common plasma
form
$\sqrt{E}\,[1+E/((\kappa_{\mathrm{p}}-3/2)T_\kappa)]^{-\kappa_{\mathrm{p}}-1}$.
The two forms differ in both exponent convention and energy
scale; a shift of the symbol $\kappa$ alone is therefore
insufficient to make them identical. Throughout this paper
$\kappa_{\mathrm{T}}=1/(q-1)$ denotes only the tail index of
Eq.~\eqref{eq:fkappa}. The values $q=1.2,\,1.4,\,1.6$
correspond to $\kappa_{\mathrm{T}}=5,\,2.5,\,1.\overline{6}$
under this stated convention and are used to span
progressively heavier tails, not as a one-to-one calibration to
a unique observational plasma-$\kappa$ parameterization. In the
Maxwellian limit $\kappa_{\mathrm{T}}\to\infty$ ($q\to 1^{+}$),
Eq.~\eqref{eq:fkappa} reduces continuously to an exponential.
The $q<1$ branch has no direct plasma-$\kappa$ analogue because
its support terminates at $E_{\max}=T/(1-q)$.

\section{Cross sections and rate integral}
\label{sec:xsec}

\subsection{Bell recommended cross section}

We adopt the analytic Bell \textit{et al.}\ representation of the
single-ionization cross section~\cite{Bell1983},
\begin{equation}
  \sigma_{\mathrm{Bell}}(E) \;=\;
  \frac{10^{-13}}{I\,E}
  \left[
    A\,\ln\!\left(\frac{E}{I}\right)
    + \sum_{i=1}^{n} B_i\left(1-\frac{I}{E}\right)^{\!i}
  \right],
  \label{eq:bell}
\end{equation}
valid for $E>I$, with $E$, $I$ in eV and $\sigma$ in cm$^{2}$.
The recommended coefficients for He\,\textsc{i}, Li\,\textsc{i},
and Be\,\textsc{i} are those of Table~5 of Bell
\textit{et~al.}~\cite{Bell1983}. We reproduce them in
Table~\ref{tab:params} and emphasise three sign choices that
must be respected to obtain the published recommended curves:
(i) $B_1,\,B_2<0$ for He\,\textsc{i}, (ii) $B_1<0$ for
Li\,\textsc{i}, and (iii) $B_1<0$ for Be\,\textsc{i}. With these
signs the He recommendation reproduces Bell's plate exactly, and
the Li, Be recommendations reproduce the tabulated coefficients
to machine precision. As Bell notes, the neutral Be\,\textsc{i}
recommendation is an \emph{empirical} estimate based on scaled
cross sections rather than on a direct set of reliable
ground-state measurements, and the He\,\textsc{i} recommendation
is accordingly the most tightly constrained of the three.

\subsection{Lotz comparison}

The classical one-shell Lotz semi-empirical
formula~\cite{Lotz1967,Lotz1968} is used as a comparator,
\begin{equation}
  \sigma_{\mathrm{Lotz}}(E) \;=\;
  \frac{a\,\zeta\,10^{-14}}{I\,E}\,
  \ln\!\left(\frac{E}{I}\right)
  \Bigl[1 - b\,\exp\!\bigl(-c\,(E/I-1)\bigr)\Bigr],
  \label{eq:lotz}
\end{equation}
where $\zeta$ is the effective number of equivalent outer-shell
electrons and $(a,b,c)$ are the semi-empirical fit parameters.
The values used here (Table~\ref{tab:params}) are the
target-specific parameters tabulated by Lotz for the neutral
species. Equation~\eqref{eq:lotz} is a one-shell reduction of
the full Lotz subshell-summed expression; we treat it strictly
as a comparator that quantifies the cross-section model
dependence of $\tau_q(T)$, not as a substitute for the
recommended Bell representation. In particular, the
high-energy tail of the total ionization cross section can in
general acquire contributions from inner subshells that the
present one-shell Lotz form omits; the Bell--Lotz differences
that we report at $E\gtrsim 10^{3}$~eV (and the corresponding
differences in $\langle\sigma v\rangle$ at the highest
temperatures) should therefore be read as a measure of model
sensitivity rather than as a definitive estimate of the
total-ionization rate.

{\color{red}%

\subsection{Position with respect to modern cross-section data}
\label{sec:uncertainty}

Since the Bell~\cite{Bell1983} recommendation (1983) and the
Lotz~\cite{Lotz1967,Lotz1968} formula (1967--68), the atomic database
landscape has advanced considerably. The most important development for
light targets is the maturation of non-perturbative close-coupling and
pseudostate calculations. For He\,\textsc{i}, the convergent close-coupling
(CCC) calculations of Fursa and Bray~\cite{FursaBray1995} and the evaluated
He data set of Ralchenko \textit{et~al.}~\cite{Ralchenko2008} provide a
modern benchmark against which the older analytic Bell/Lotz forms can be
checked. For Li\,\textsc{i}, the total-ionization comparison summarized by
Bray, Beck and Plottke~\cite{BrayLi1999} includes a Bray CCC curve and the
available experimental total-ionization data. For Be\,\textsc{i}, the
BSR/CCC study of Zatsarinny, Bartschat, Fursa and Bray~\cite{ZatsarinnyBe2016}
was designed precisely to provide reliable neutral-Be scattering and
ionization data for modelling applications. Analytic models have also
advanced: the binary-encounter-Bethe (BEB) model of Kim and
Rudd~\cite{KimRudd1994} reproduces measured total single-ionization cross
sections of many neutrals without target-specific fitting. Recommended rate
compilations for solar and fusion modelling---CHIANTI~\cite{DelZanna2022,Dere2007},
the ADAS family and the evaluated sets of Hahn and Savin~\cite{Hahn2017}
and Bryans \textit{et~al.}~\cite{Bryans2006}---therefore no longer rest only
on the older analytic formulae.

{\color{red}
The calculation also carries a modern benchmark curve through the same rate
integral for \emph{each} of the three neutrals. For He the curve represents
the Ralchenko \textit{et~al.} recommended helium benchmark. For Li and Be,
the benchmark points are approximate manual digitizations of the
total-ionization curves shown in Fig.~1 of Ref.~\cite{BrayLi1999} and
Fig.~6 of Ref.~\cite{ZatsarinnyBe2016}, respectively. Only points within
the published plotting range up to $100$~eV are stored. Graphical
extraction and light smoothing imply an estimated $5$--$10\%$ uncertainty,
so differences below that level should not be overinterpreted. These curves
are sensitivity references rather than replacements for the original
authors' numerical tables. The same interpolation and quadrature procedure
is applied to Bell, Lotz, and the benchmark input on equal footing.}

For a benchmark table with points $(E_j,\sigma_j)$ the interpolation used in
the code is log--log interpolation over the positive cross-section range,
\begin{equation}
  \sigma_{\rm bench}(E)=
  \exp\!\left[\operatorname{interp}\bigl(\ln E;
  \ln E_j,\ln\sigma_j\bigr)\right],
  \label{eq:benchmark-interp}
\end{equation}
with a linear near-threshold onset and $\sigma_{\rm bench}(E)=0$ below the
first ionization threshold. Above the last benchmark point a smooth log--log
continuation fixed by the last two tabulated points is used; this affects
mostly the far high-energy tail of the hottest stress-test cases and is
recorded explicitly in the reproducibility script. Thus the strict
cross-section and rate comparisons are Bell--Lotz--modern-benchmark for
He\,\textsc{i}, Li\,\textsc{i}, and Be\,\textsc{i}.

The effect of the modern benchmark is deliberately shown rather than
assumed. In Fig.~\ref{fig:xsec}, the He benchmark lies almost on the same
envelope as Bell and Lotz near the peak, the Li benchmark stays between the
Bell and Lotz rates over much of the Maxwellian range, and the Be BSR/CCC
benchmark is visibly above both analytic curves around and above the peak.
In the strict Maxwellian limit, Table~\ref{tab:ratediff-strict} shows that
replacing Bell by the modern benchmark changes the rate by $+0.7\%$, $-0.7\%$,
and $-3.8\%$ for He at $T=10,100,1000$~eV; by $-17.2\%$, $+6.7\%$, and
$+31.7\%$ for Li; and by $-14.7\%$, $+30.5\%$, and $+118.9\%$ for Be. The
large high-temperature Be difference is extrapolation-sensitive because the
digitized source curve ends at $100$~eV; it is an uncertainty flag, not a
precision benchmark. The physics conclusion is therefore more nuanced: the
large low-temperature enhancement generated by a super-extensive EEDF is not an
artifact of using only Bell/Lotz, but the absolute Be rate above the peak is
more sensitive to the chosen modern atomic data set than the He rate. In the
single-Maxwellian limit our He\,\textsc{i} Bell rate is also cross-checked
against the Voronov analytical fit~\cite{Voronov1997}; that comparison is
carried out in Sec.~\ref{sec:numerics} and Table~\ref{tab:validation-voronov}.
}

\begin{table*}[t]
\centering
\caption{Parameters used in Eqs.~\eqref{eq:bell}--\eqref{eq:lotz}.
Bell coefficients are from Table~5 of
Ref.~\cite{Bell1983}. Lotz one-shell parameters are those of
Refs.~\cite{Lotz1967,Lotz1968} for the neutral targets.
Ionization thresholds $I$ are NIST recommended values from the
Atomic Spectra Database~\cite{NIST_ASD}.}
\label{tab:params}
\setlength{\tabcolsep}{6pt}
\begin{ruledtabular}
\begin{tabular}{l d d d d d d d c d d d}
\multicolumn{1}{c}{target} &
\multicolumn{1}{c}{$I$\,(eV)} &
\multicolumn{1}{c}{$A$} &
\multicolumn{1}{c}{$B_1$} &
\multicolumn{1}{c}{$B_2$} &
\multicolumn{1}{c}{$B_3$} &
\multicolumn{1}{c}{$B_4$} &
\multicolumn{1}{c}{$B_5$} &
$\zeta$ &
\multicolumn{1}{c}{$a$} &
\multicolumn{1}{c}{$b$} &
\multicolumn{1}{c}{$c$} \\ \hline
He\,\textsc{i} & 24.587 & 0.5720 & -0.3440 & -0.5230 & 3.4450 & -6.8210 & 5.5780 & 2 & 4.00 & 0.75 & 0.46 \\
Li\,\textsc{i} &  5.392 & 0.0854 & -0.0040 &  0.7573 & -0.1779 & \multicolumn{1}{c}{---} & \multicolumn{1}{c}{---} & 1 & 4.00 & 0.70 & 0.30 \\
Be\,\textsc{i} &  9.323 & 0.9239 & -0.7697 &  0.3619 & \multicolumn{1}{c}{---} & \multicolumn{1}{c}{---} & \multicolumn{1}{c}{---} & 2 & 4.00 & 0.70 & 0.30 \\
\end{tabular}
\end{ruledtabular}
\end{table*}

\subsection{Rate integral}

The Tsallis rate coefficient for an impact-ionization event
$\mathrm{X}+e^{-}\to\mathrm{X}^{+}+2e^{-}$ is
\begin{equation}
  \tau_q(T) \;=\;
  \int_{I}^{E_{\mathrm{up}}} v(E)\,\sigma(E)\,
  f_q^{\mathrm{NM}}(E)\,dE,
  \label{eq:rate}
\end{equation}
with $v(E)=100\sqrt{2eE/m_e}$ in cm\,s$^{-1}$ when $E$ is in eV.
For $q<1$ we use the exact mixture cutoff
$E_{\mathrm{up}}=T_{\mathrm{hot}}/(1-q)$; for $q\ge 1$ the released
calculation uses $E_{\mathrm{up}}=\max(80\,T_{\mathrm{hot}},\,50 I)$.
The residual finite-cutoff sensitivity of the algebraic tail is
quantified in Sec.~\ref{sec:cutoff}.
We emphasise that the symbol $\tau_q(T)$ used throughout this
paper denotes a \emph{rate coefficient}---i.e.\ the EEDF-averaged
product $\langle \sigma v\rangle_q$, with units of
cm$^{3}$\,s$^{-1}$---and not a time or lifetime, despite the
common use of $\tau$ for the latter in other contexts. The
notation is retained for compactness and continuity with our
earlier work~\cite{Khalfaoui2022}, but readers familiar with the
alternative notation may equivalently set
$\tau_q(T)\equiv\langle\sigma v\rangle_q(T)$ in every formula
and figure caption below.

\subsection{Numerical method and validation}
\label{sec:numerics}

\paragraph*{Method.}
\RevI{The reader interested only in the physical results may skip to
Sec.~\ref{sec:maxwellian}; this subsection documents the
numerical scheme purely so that every number in the paper can be
regenerated from the released code.} The rate integral
Eq.~\eqref{eq:rate} is a one-dimensional
integral of a smooth integrand on a finite or semi-infinite
interval. We evaluate it by a fixed-order composite
Gauss--Legendre quadrature, which is the optimal method for
smooth integrands of this type because the error decays
exponentially with the number of nodes for analytic
integrands~\cite{DavisRabinowitz1984,GolubWelsch1969,Trefethen2008}.
Concretely, on an interval $[a,b]$ the rule reads
\begin{equation}
  \int_{a}^{b}\!g(E)\,dE
  \;\approx\; \frac{b-a}{2}\sum_{k=1}^{N}\,w_{k}\,
  g\!\left(\frac{b-a}{2}\,x_{k}+\frac{b+a}{2}\right),
  \label{eq:GL}
\end{equation}
where $\{x_{k},w_{k}\}_{k=1}^{N}$ are the nodes and weights of
the $N$-point Gauss--Legendre rule on $[-1,1]$, computed once
from the eigendecomposition of the Jacobi matrix associated
with the Legendre polynomials by the standard Golub--Welsch
algorithm~\cite{GolubWelsch1969} as implemented in
\textsc{numpy}~\cite{Harris2020}. We use $N=96$ nodes per
sub-interval throughout. The composite version of
Eq.~\eqref{eq:GL} that we use splits the energy range into a
near-threshold sub-interval $[I,E_{\rm mid}]$, with
$E_{\rm mid}=\max(3I,\,5T_{\rm hot})$, and a far sub-interval
$[E_{\rm mid},E_{\mathrm{up}}]$. The split is essential at low
$T$: when $T\ll I$, the integrand is very sharply peaked just
above threshold, and a single uniform quadrature rule on the
full range underestimates the peak density of quadrature points
and therefore the integral. The same idea underlies the
adaptive subdivision strategies of standard packages such as
QUADPACK~\cite{Piessens1983}.

The Python implementation follows the rate formula literally.
For each species $s$, atomic-data input $m$, temperature $T_i$,
Tsallis index $q$, and hot-electron fraction $f_{\mathrm{hot}}$,
it evaluates
\begin{equation}
\begin{split}
  \tau_{s,m}(T_i,q,f_{\mathrm{hot}})=
  &\sum_{r=\mathrm{bulk,hot}} w_r\,100
  \sqrt{\frac{2e}{m_e}}\\
  &\times\int_{I_s}^{E_{\mathrm{up}}}
  \sigma_{s,m}(E)\,E^{1/2} f_q(E;T_r)\,dE ,
\end{split}
  \label{eq:python-rate}
\end{equation}
with $(w_{\mathrm{bulk}},w_{\mathrm{hot}})=(1-f_{\mathrm{hot}},
f_{\mathrm{hot}})$ and $T_{\mathrm{hot}}=10T_{\mathrm{bulk}}$.
For $q<1$ the global upper limit is the hot-component cutoff; the bulk
component vanishes above its own smaller cutoff. For $q\ge 1$ the finite
limit stated after Eq.~\eqref{eq:rate} is used. The script caches repeated
parameter combinations, evaluates each integrand as a NumPy array on the
96 Gauss--Legendre nodes of each sub-interval, and forms the weighted sum
with vector operations. Gamma functions are evaluated with
SciPy~\cite{Virtanen2020}, and figures are generated with
Matplotlib~\cite{Hunter2007}. The explicit factor $100\sqrt{2e/m_e}$
converts the speed to cm\,s$^{-1}$, giving rates in cm$^3$\,s$^{-1}$.

\paragraph*{Validation.}
We checked the composite Gauss--Legendre integrator
independently against adaptive Gauss--Kronrod quadrature
(QUADPACK) using \texttt{scipy.integrate.quad} with tight
relative tolerance. The released file
\texttt{quadrature\_validation.csv} contains 27 representative
checks covering all species, $T=1,10,100$~eV, and the
Maxwellian, $q=1.2$, and $q=1.4$ cases. The largest relative
difference is below $7\times 10^{-4}\%$. As a further consistency check,
Table~\ref{tab:validation} lists the strict-Maxwellian
($q=1,\,f_{\mathrm{hot}}=0$) Bell-based rates at three
representative temperatures and compares them with the values
obtained independently from the earlier code of
Ref.~\cite{Khalfaoui2022}. Agreement is better than $0.3\%$ for
every entry.

We also benchmarked the Maxwellian limit of our code against an
external, widely used compilation that does not share
implementation details with either of the codes in
Table~\ref{tab:validation}. Voronov's analytical
fit~\cite{Voronov1997} reproduces the Belfast group's
recommended thermal rates to within a few percent for every
neutral and ion from H to Ni; we use it in the He~I form
$\langle\sigma v\rangle=A\,(1+P\sqrt{U})\,(X+U)^{-1}\,U^K
e^{-U}$ with $U\equiv I_p/k_B T_{\mathrm{bulk}}$, $I_p=24.6$~eV,
$A=1.75\times 10^{-8}$~cm$^3$\,s$^{-1}$, $P=0$, $X=0.18$,
$K=0.35$. Table~\ref{tab:validation-voronov} compares the He~I
strict-Maxwellian rates of the present work with the Voronov
fit at the same three temperatures of
Table~\ref{tab:validation}. The agreement is below $1\%$ over
two decades of temperature, well within Voronov's quoted
fit accuracy. The Voronov comparison is used only as an
independent He~I check here; no claim is made for Li~I or
Be~I unless their species-specific coefficients are
implemented explicitly.

\begin{table*}[t]
\centering
\caption{Validation of the numerical rate-integral scheme.
Listed are strict Maxwellian Bell-based rate coefficients
$\tau_M$ in units of $10^{-8}$~cm$^{3}$\,s$^{-1}$ at three
representative bulk temperatures, computed with the present
96-node composite Gauss--Legendre integrator (\textit{this work},
denoted T.W.) and with the independent code of
Ref.~\cite{Khalfaoui2022} (denoted Ref.). The relative
difference $\Delta\equiv 100\,(\tau_{\rm T.W.}
-\tau_{\rm Ref.})/\tau_{\rm Ref.}$ never exceeds $0.3\%$.}
\label{tab:validation}
\begin{ruledtabular}
\begin{tabular}{r c c c c c c c c c}
 & \multicolumn{3}{c}{He} & \multicolumn{3}{c}{Li} & \multicolumn{3}{c}{Be}\\
\cline{2-4}\cline{5-7}\cline{8-10}
$T$\,(eV) & T.W. & Ref. & $\Delta$\,(\%)
          & T.W. & Ref. & $\Delta$\,(\%)
          & T.W. & Ref. & $\Delta$\,(\%) \\
\hline
$10$    & $0.07695$ & $0.07677$ & $+0.24$
        & $6.486$   & $6.491$   & $-0.07$
        & $3.014$   & $3.016$   & $-0.08$ \\
$100$   & $1.983$   & $1.981$   & $+0.11$
        & $7.916$   & $7.920$   & $-0.05$
        & $10.08$   & $10.09$   & $-0.09$ \\
$1000$  & $2.297$   & $2.295$   & $+0.10$
        & $3.716$   & $3.722$   & $-0.16$
        & $7.708$   & $7.699$   & $+0.12$ \\
\end{tabular}
\end{ruledtabular}
\end{table*}

\begin{table}[t]
\centering
\caption{External Maxwellian benchmark: He~I rate coefficient
$\tau_M$ (units $10^{-8}$~cm$^{3}$\,s$^{-1}$) at three
temperatures, computed with the present integrator
($\tau_{\rm T.W.}$) and with the analytical fit of
Voronov~\cite{Voronov1997}
($\tau_{\rm Vor.}$). Relative difference $\Delta\equiv
100\,(\tau_{\rm T.W.}-\tau_{\rm Vor.})/\tau_{\rm Vor.}$
sits below~$1\%$, well within Voronov's quoted few-percent
fit accuracy. Equivalent results for Li~I and Be~I are
included in the public reproducibility package.}
\label{tab:validation-voronov}
\begin{ruledtabular}
\begin{tabular}{r c c c}
$T$\,(eV) & $\tau_{\rm T.W.}$ & $\tau_{\rm Vor.}$ & $\Delta$\,(\%) \\
\hline
$10$    & $0.07695$ & $0.07761$ & $-0.85$ \\
$100$   & $1.983$   & $1.966$   & $+0.86$ \\
$1000$  & $2.297$   & $2.282$   & $+0.66$ \\
\end{tabular}
\end{ruledtabular}
\end{table}

\subsection{Cutoff convergence}
\label{sec:cutoff}

For $q\ge 1$ the formal upper limit is infinite. The released
figures use $E_{\mathrm{up}}=\max(80\,T_{\mathrm{hot}},\,50 I)$,
which is ample for the Maxwellian and for the moderate $q=1.2$
tail but can leave a visible remainder for heavier tails. We
therefore repeated the Bell calculation with adaptive QUADPACK
integration and increased the multiplier from 80 to 2000 at
$T_{\mathrm{bulk}}=1$ and $10$~eV. The relative difference
$(\tau_{80}-\tau_{2000})/\tau_{2000}$ is reported in
Table~\ref{tab:cutoff} and in the released
\texttt{cutoff\_convergence.csv}.

\begin{table}[t]
\centering
\caption{Finite-cutoff test for the super-extensive branch.
Listed is $100(\tau_{80}-\tau_{2000})/\tau_{2000}$ for Bell
cross sections at $f_{\mathrm{hot}}=0.10$, where the subscripts
denote the multiplier of $T_{\mathrm{hot}}$ in the upper
cutoff. The $q=1.2$ error is below $0.015\%$ in every tested
case. At $q=1.4$ it remains below $3.8\%$; $q=1.6$ is
substantially more cutoff-sensitive and is retained only as a
qualitative heavy-tail stress test.}
\label{tab:cutoff}
\begin{ruledtabular}
\begin{tabular}{l c c c}
species & $q$ & $T_{\mathrm{bulk}}$\,(eV) & rel.\ change \\ \hline
He & $1.4$ & $1$  & $-3.76\%$  \\
He & $1.4$ & $10$ & $-0.65\%$  \\
He & $1.6$ & $1$  & $-25.43\%$ \\
He & $1.6$ & $10$ & $-10.80\%$ \\
Be & $1.4$ & $1$  & $-2.35\%$  \\
Be & $1.4$ & $10$ & $-0.35\%$  \\
Be & $1.6$ & $1$  & $-21.86\%$ \\
Be & $1.6$ & $10$ & $-8.57\%$  \\
Li & $1.4$ & $1$  & $-0.66\%$  \\
Li & $1.4$ & $10$ & $-0.12\%$  \\
Li & $1.6$ & $1$  & $-9.94\%$  \\
Li & $1.6$ & $10$ & $-4.02\%$  \\
\end{tabular}
\end{ruledtabular}
\end{table}

\section{Cross-section comparison and $q=1$ benchmarks}
\label{sec:maxwellian}

\subsection{Bell vs.\ Lotz cross sections}

\begin{figure*}[t]
\centering
\includegraphics[width=0.95\linewidth]{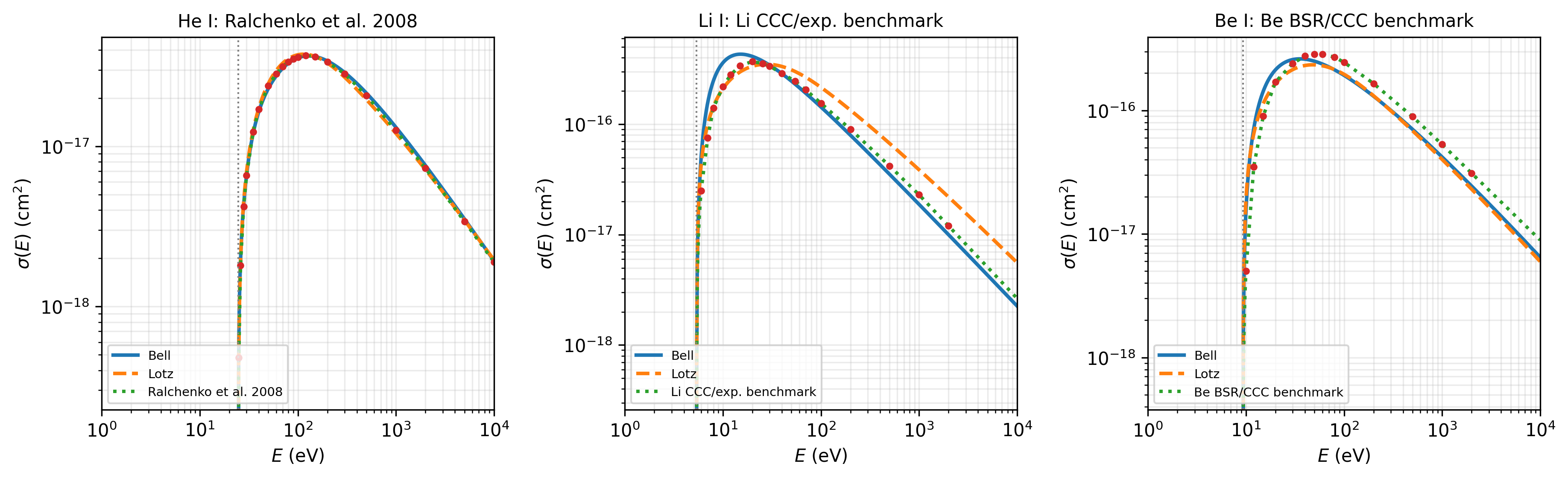}
\caption{Single-ionization cross sections of neutral He (left),
Li (centre), and Be (right) over $E\in[1,10^4]$~eV. Solid blue:
Bell representation Eq.~\eqref{eq:bell} with the coefficients of
Table~\ref{tab:params}. Dashed orange: one-shell Lotz form
Eq.~\eqref{eq:lotz}. Green dotted lines and markers: modern benchmark
curves: Ralchenko \textit{et~al.}~\cite{Ralchenko2008} for He\,\textsc{i},
the CCC/experimental Li\,\textsc{i} benchmark of
Refs.~\cite{BrayLi1999,BrayLiLike1995}, and the Be\,\textsc{i} BSR/CCC benchmark of Ref.~\cite{ZatsarinnyBe2016}. Thus, the
label ``Ralchenko \textit{et~al.}'' applies only to the He panel, not to Li or Be. Dotted vertical lines:
first-ionization thresholds. Detailed discussion in Sec.~\ref{sec:maxwellian}.}
\label{fig:xsec}
\end{figure*}

Figure~\ref{fig:xsec} compares Bell Eq.~\eqref{eq:bell} and Lotz
Eq.~\eqref{eq:lotz} for the three neutrals. Three species-specific
patterns emerge. For He, the two formulas are nearly
indistinguishable over the entire energy range: both predict the
same threshold onset, the same peak near
$E\sim 100$~eV, and the same high-energy decay. For Li, the Bell
curve peaks slightly earlier and decays faster than the Lotz
curve, so that Lotz assigns more weight to moderate and high
energies. Be is intermediate: the two curves coincide near the
peak but differ at high energy. This ranking---He $\ll$ Be
$\ll$ Li in model sensitivity---is the first principal result of
the paper. Bell's own
discussion~\cite{Bell1983} supports this ordering: the He\,\textsc{i}
recommendation is based on direct experimental and theoretical
information, the Be\,\textsc{i} recommendation is an empirical
estimate from scaled cross sections, and the neutral-Li
cross-section data at moderate and high energy are likewise less
tightly constrained.

\subsection{Strict Maxwellian benchmark $(q{=}1,\,f_{\mathrm{hot}}{=}0)$}

\begin{figure*}[t]
\centering
\includegraphics[width=0.95\linewidth]{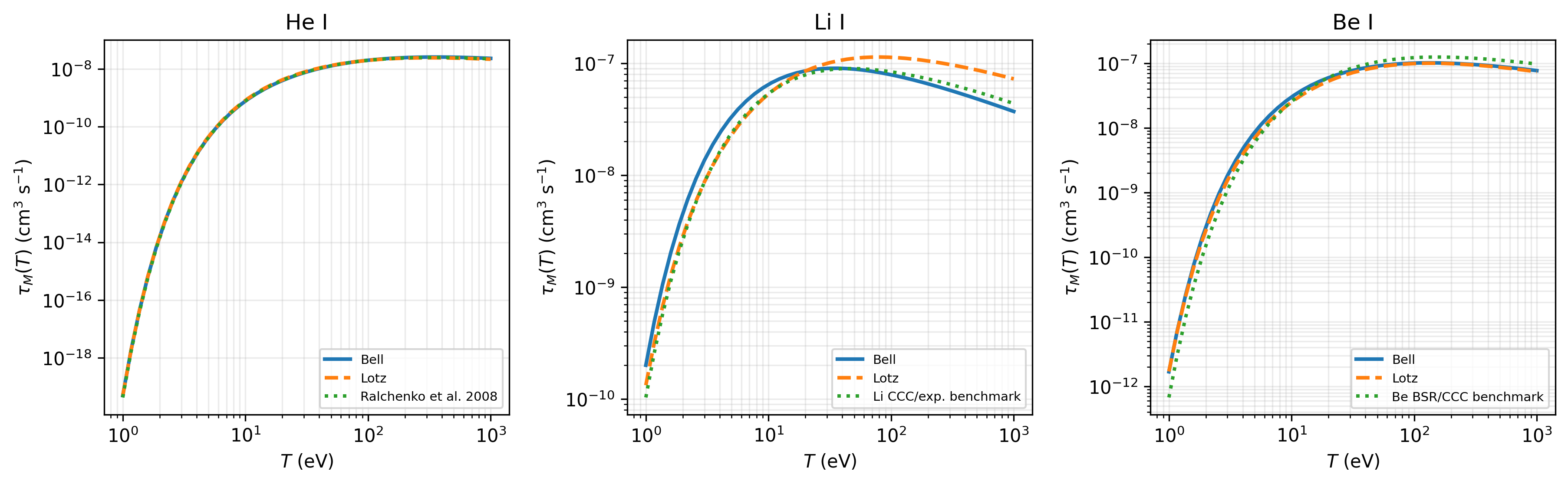}
\caption{Strict Maxwellian rate coefficients
$\tau_M(T)\equiv\tau_q(T)|_{q=1,f_{\mathrm{hot}}=0}$ for He
(left), Li (centre) and Be (right) on log-log axes,
$T\in[1,10^3]$~eV. Solid blue: Bell. Dashed orange: Lotz. Green dotted curves are the rates obtained by inserting the corresponding
modern benchmark data into the same EEDF integral. In this $(q,f_{\mathrm{hot}})=(1,0)$ limit the
two-temperature mixture Eq.~\eqref{eq:fNM} collapses to a single
Maxwellian. Quantitative differences at $T=10$, $100$, $1000$~eV are
listed in Table~\ref{tab:ratediff-strict}.}
\label{fig:maxwellian-strict}
\end{figure*}

Figure~\ref{fig:maxwellian-strict} shows the strict Maxwellian
rate coefficient obtained by convolving Bell, Lotz, and the
species-resolved modern benchmark cross sections with a single-temperature
$f_M(E;T)$. This is the
correct limit in which the present calculation can be
positioned with respect to standard Maxwellian rate
tables~\cite{Kato1991,ArnaudRothenflug1985}, because only here
does Eq.~\eqref{eq:fNM} reduce to a pure single-temperature
thermal plasma; we do not, however, perform a detailed
numerical comparison with those compilations, which would
require a careful matching of the underlying cross-section
data and is beyond the scope of the present paper. The
species hierarchy found in Fig.~\ref{fig:xsec} is transmitted
to the rate coefficients: Bell, Lotz, and the modern benchmark agree closely
for He, while the Li and Be panels expose the remaining species-dependent
atomic-data sensitivity.
Table~\ref{tab:ratediff-strict} quantifies the agreement at
three temperatures. The He benchmark remains within the Bell--Lotz envelope
at the few-percent level; Li shows a modest benchmark offset compared with
the much larger Lotz high-temperature tail; and Be exhibits a larger
modern-benchmark shift above the peak (about $21$--$26\%$ at
$100$--$1000$~eV). The very large Li Bell--Lotz difference at high $T$ reflects
the fact that the rate integral at $T=1$~keV samples the Lotz
tail, which lies a factor of a few above the Bell tail in
Fig.~\ref{fig:xsec}.

\begin{table*}[t]
\centering
\caption{Representative strict-Maxwellian rate-coefficient comparison at
$(q,f_{\mathrm{hot}})=(1,0)$. The modern benchmark column uses Ralchenko
\textit{et~al.}~\cite{Ralchenko2008} for He\,\textsc{i}, the Li\,\textsc{i}
CCC/experimental benchmark of Refs.~\cite{BrayLi1999,BrayLiLike1995}, and
the Be\,\textsc{i} BSR/CCC benchmark of Ref.~\cite{ZatsarinnyBe2016}. The
relative differences are computed with respect to Bell. All rates are in
cm$^3$\,s$^{-1}$.}
\label{tab:ratediff-strict}
\begin{ruledtabular}
\begin{tabular}{l c c c c c}
target & $T$\,(eV) & Bell & Lotz & modern benchmark & rel. diff. range \\ \hline
He & $10$   & $7.70\times 10^{-10}$ & $8.14\times 10^{-10}$ & $7.75\times 10^{-10}$ & $+0.7\%$ to $+5.8\%$ \\
He & $100$  & $1.98\times 10^{-8}$  & $1.97\times 10^{-8}$  & $1.97\times 10^{-8}$  & $-0.7\%$ to $-0.7\%$ \\
He & $1000$ & $2.30\times 10^{-8}$  & $2.15\times 10^{-8}$  & $2.21\times 10^{-8}$  & $-6.6\%$ to $-3.9\%$ \\
Li & $10$   & $6.49\times 10^{-8}$  & $5.32\times 10^{-8}$  & $5.37\times 10^{-8}$  & $-17.9\%$ to $-17.1\%$ \\
Li & $100$  & $7.92\times 10^{-8}$  & $1.13\times 10^{-7}$  & $8.45\times 10^{-8}$  & $+6.7\%$ to $+43.2\%$ \\
Li & $1000$ & $3.72\times 10^{-8}$  & $7.28\times 10^{-8}$  & $4.89\times 10^{-8}$  & $+31.7\%$ to $+95.9\%$ \\
Be & $10$   & $3.01\times 10^{-8}$  & $2.51\times 10^{-8}$  & $2.57\times 10^{-8}$  & $-16.6\%$ to $-14.7\%$ \\
Be & $100$  & $1.01\times 10^{-7}$  & $9.92\times 10^{-8}$  & $1.32\times 10^{-7}$  & $-1.6\%$ to $+30.5\%$ \\
Be & $1000$ & $7.71\times 10^{-8}$  & $7.41\times 10^{-8}$  & $1.69\times 10^{-7}$  & $-3.8\%$ to $+118.9\%$ \\
\end{tabular}
\end{ruledtabular}
\end{table*}

\subsection{Two-temperature mixture at $q=1$}

\begin{figure*}[t]
\centering
\includegraphics[width=0.95\linewidth]{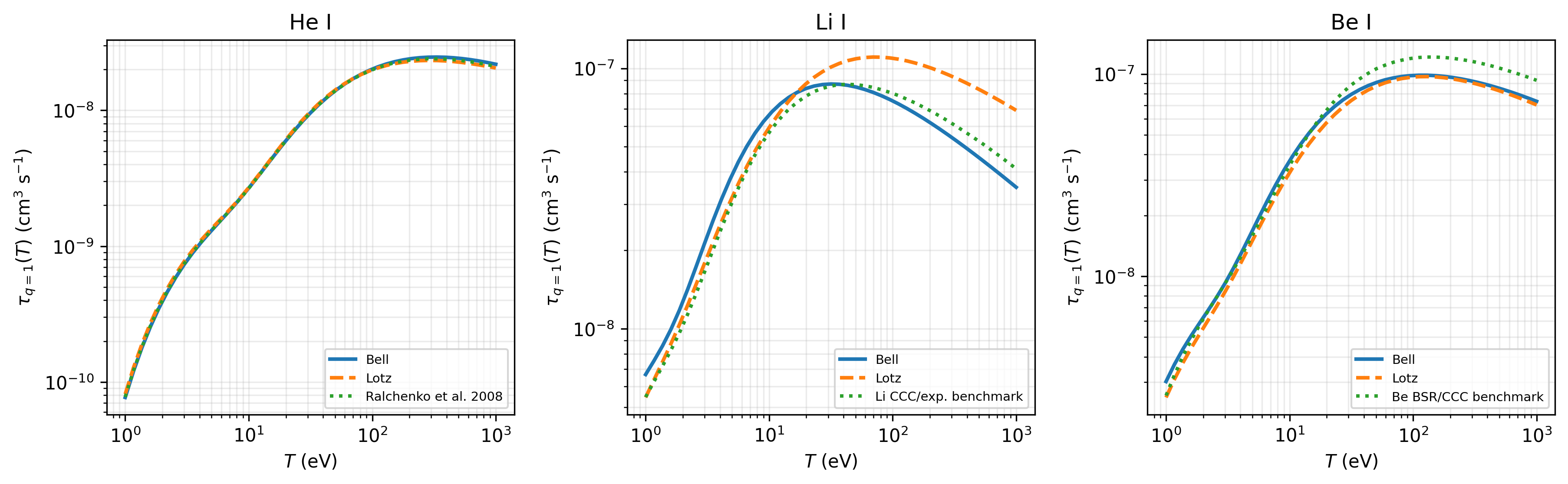}
\caption{Two-temperature rate coefficient $\tau_q(T)$ at $q=1$,
$f_{\mathrm{hot}}=0.10$, $T_{\mathrm{hot}}=10\,T_{\mathrm{bulk}}$.
Solid blue: Bell. Dashed orange: Lotz. Green dotted curves use the
species-resolved modern benchmarks in the same rate integral: Ralchenko
\textit{et~al.}~\cite{Ralchenko2008} for He\,\textsc{i}, the Li CCC
benchmark, and the Be BSR/CCC benchmark. Compared with
Fig.~\ref{fig:maxwellian-strict}, the $10\%$ hot population raises the
curves at low $T_{\mathrm{bulk}}$ while leaving the high-$T$ plateau
unchanged.}
\label{fig:twoT-q1}
\end{figure*}

Figure~\ref{fig:twoT-q1} shows the rate coefficient at $q=1$
with $f_{\mathrm{hot}}=0.10$. Because the hot admixture is
retained, this is \emph{not} a single-Maxwellian benchmark; it
is a two-temperature mixture in which each branch is separately
Maxwellian. Numerically, the $10\%$ hot fraction mostly raises
the low-$T$ wing of the curves (via hot-electron injection above
threshold), while the high-$T$ plateau is almost unchanged. The
Bell--Lotz--benchmark pattern is essentially the one of
Fig.~\ref{fig:maxwellian-strict}: agreement within a few percent
for He, moderate differences for Be, and substantial differences
for Li at high $T$.

\section{Non-Maxwellian Tsallis results}
\label{sec:tsallis-results}

This section inserts all three atomic-data inputs--Bell, Lotz,
and the modern benchmark curve--into the full two-temperature Tsallis
rate integral.  For He the benchmark is the Ralchenko
\textit{et~al.} data set; for Li and Be it is the corresponding
CCC or BSR/CCC benchmark introduced above.  Representative results at
four values of $q$ are shown in Figs.~\ref{fig:q1}--\ref{fig:q14}.
For each $q$ the colour identifies the hot-electron fraction
$f_{\mathrm{hot}}\in\{0,0.01,0.06,0.10,0.30,0.40\}$.  The
figures separate the atomic-data input by columns rather than by line style:
Bell, Lotz, and the modern benchmark are displayed side by side.  The modern
benchmark column uses Ralchenko \textit{et~al.} only for He; Li and Be use
their corresponding CCC or BSR/CCC benchmark curves.  Figure~\ref{fig:qgt1}
summarises the super-extensive regime $q>1$ with the same split-panel
comparison.

\subsection{Maxwellian baseline $q=1$}

\begin{figure*}[t]
\centering
\includegraphics[width=0.95\linewidth]{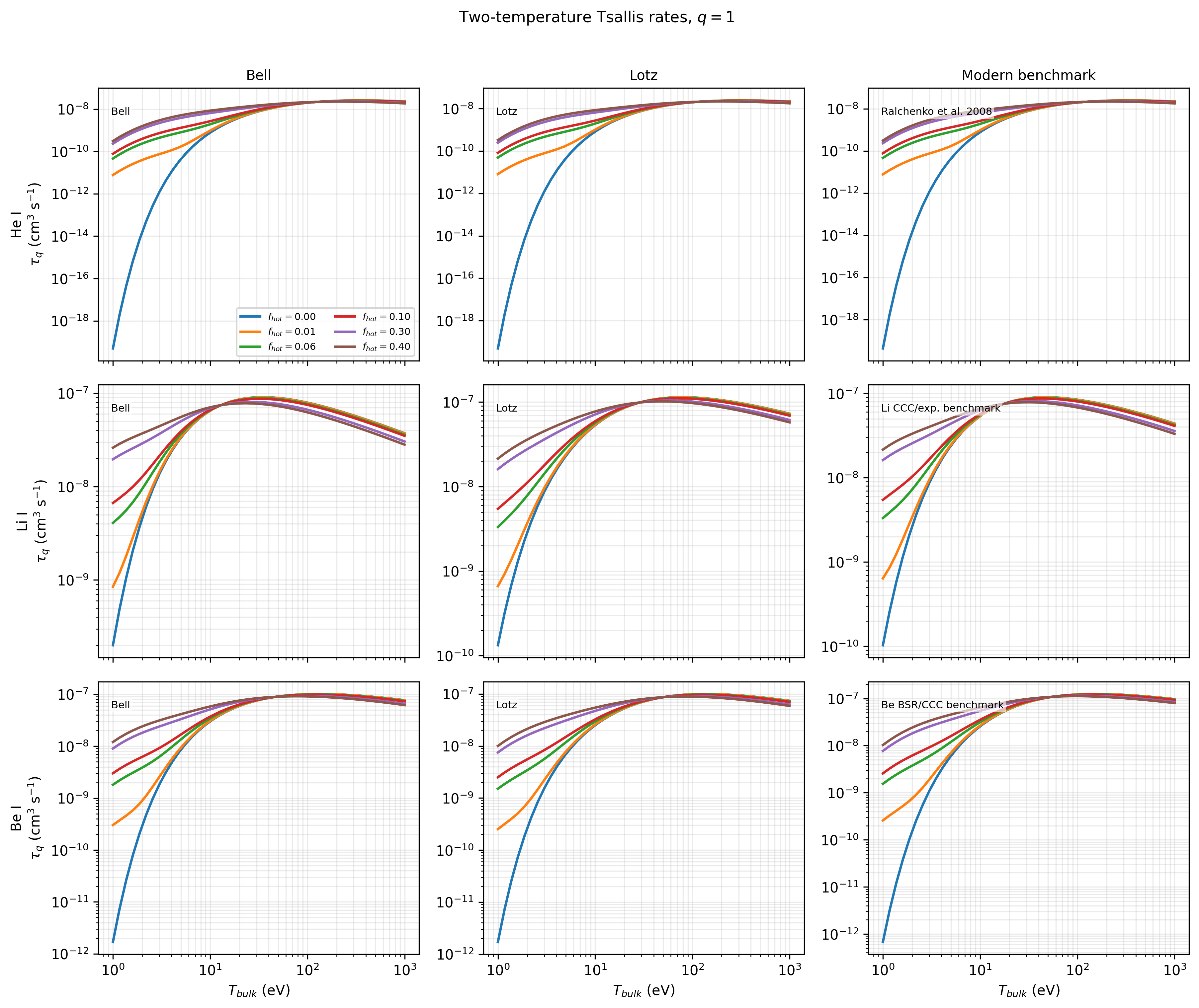}
\caption{Rate coefficients at $q=1$ vs.\ bulk temperature
$T_{\mathrm{bulk}}$.  Rows correspond to He, Li, and Be; columns correspond to
Bell, Lotz, and the modern benchmark input. Colours label
$f_{\mathrm{hot}}=0,0.01,0.06,0.10,0.30,0.40$.  Only the He benchmark column
uses the Ralchenko \textit{et~al.} evaluated data; the Li and Be benchmark
columns use the corresponding CCC/experimental and BSR/CCC benchmark curves,
respectively.  The low-$T$ spread is many orders of magnitude for He, less for
Be, and a factor $\lesssim 10$ for Li, reflecting
$I_p/k_BT_{\mathrm{bulk}}$.}
\label{fig:q1}
\end{figure*}

Figure~\ref{fig:q1} is the $q=1$ Tsallis family for each
species and for each atomic-data input. The rate coefficients rise steeply
above threshold, reach a broad maximum, and decay slowly at high temperature.
Within every panel, the curve with $f_{\mathrm{hot}}=0$ is the single-
temperature Maxwellian reference; at the lowest bulk temperatures, even a
$1\%$ hot fraction adds more above-threshold flux than the bulk Maxwellian
supplies by itself, and the rate increases with $f_{\mathrm{hot}}$. This
effect is most visible for He, whose higher ionization threshold makes the
bulk contribution negligible at $T_{\mathrm{bulk}}\sim 1$--$10$~eV. For Li,
whose threshold is smaller than the low-temperature bulk by factor of a few,
the bulk itself already supplies appreciable above-threshold flux at all
plotted temperatures and the spread in $f_{\mathrm{hot}}$ is correspondingly
smaller. Be is intermediate.

\subsection{Sub-extensive regime $q<1$}

\begin{figure*}[t]
\centering
\includegraphics[width=0.95\linewidth]{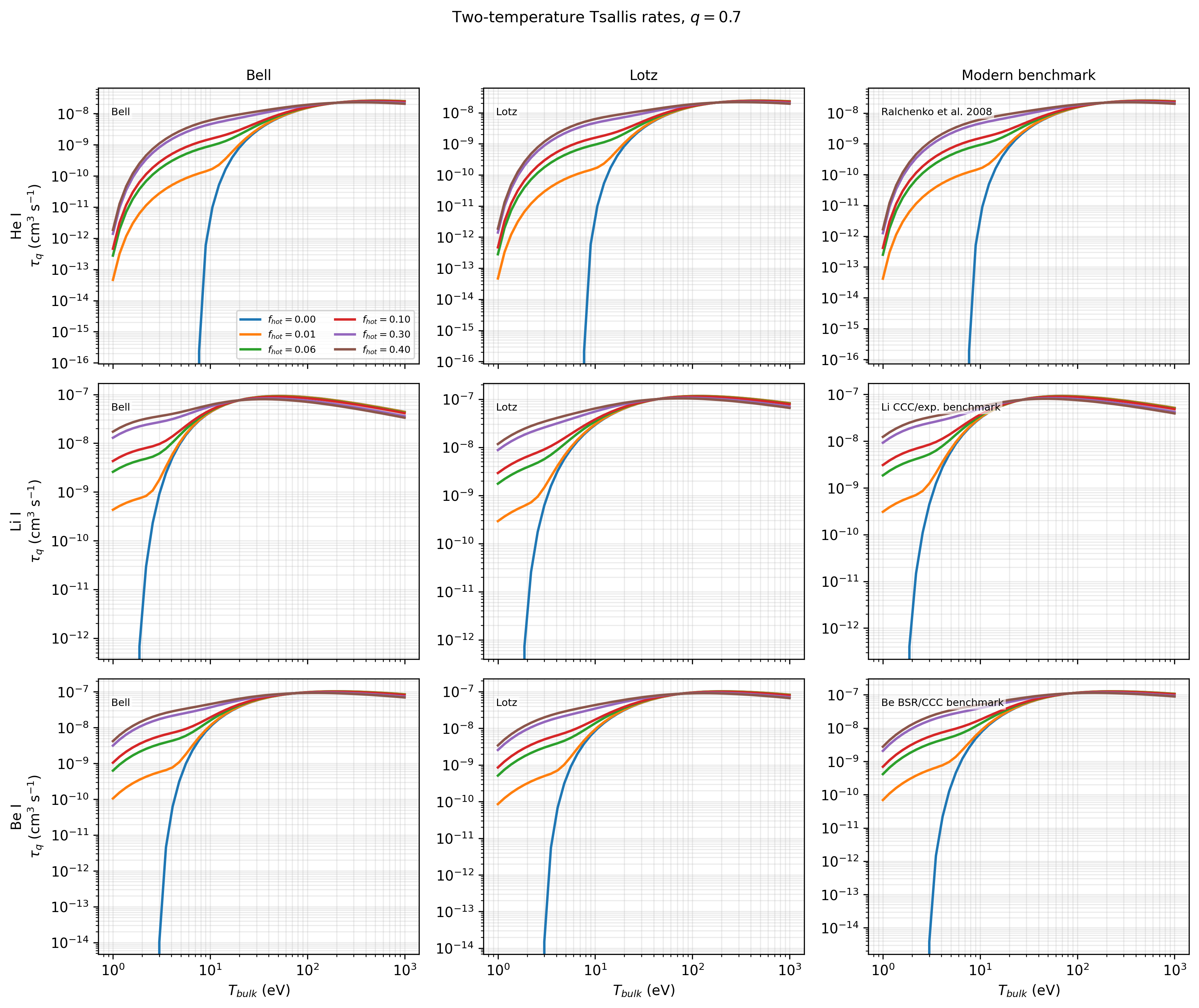}
\caption{Rate coefficients at $q=0.7$.  Rows are He, Li, and Be;
columns are Bell, Lotz, and modern benchmark input, with the Ralchenko
\textit{et~al.} data used only in the He benchmark panel.  Colours label
$f_{\mathrm{hot}}$. The EEDF is truncated at
$E_{\max}(T)=T/(1-q)\simeq 3.3\,T$, giving a hard tail cut-off.
Compared with Fig.~\ref{fig:q1}, $\tau_q$ is most depressed for He at low
$T_{\mathrm{bulk}}$, mildly for Be, and barely for Li. Increasing
$f_{\mathrm{hot}}$ partially compensates the depletion, while the three
columns display the residual atomic-data sensitivity.}
\label{fig:q07}
\end{figure*}

When $q<1$ the EEDF acquires a hard cut-off at
$E_{\max}(T)=T/(1-q)$ and the high-energy tail is depleted.
Figure~\ref{fig:q07} shows that this depletion is already
important at $q=0.7$. The suppression is species-dependent: it
is strongest for He, for which the rate integral relies heavily
on the suprathermal part of the EEDF, and weakest for Li, which
can ionize efficiently even after mild tail removal. Be again
lies in between. The spread of the curves with
$f_{\mathrm{hot}}$ indicates that the hot component partially
compensates the tail depletion, but the systematic suppression
caused by $q<1$ persists.

\begin{figure*}[t]
\centering
\includegraphics[width=0.95\linewidth]{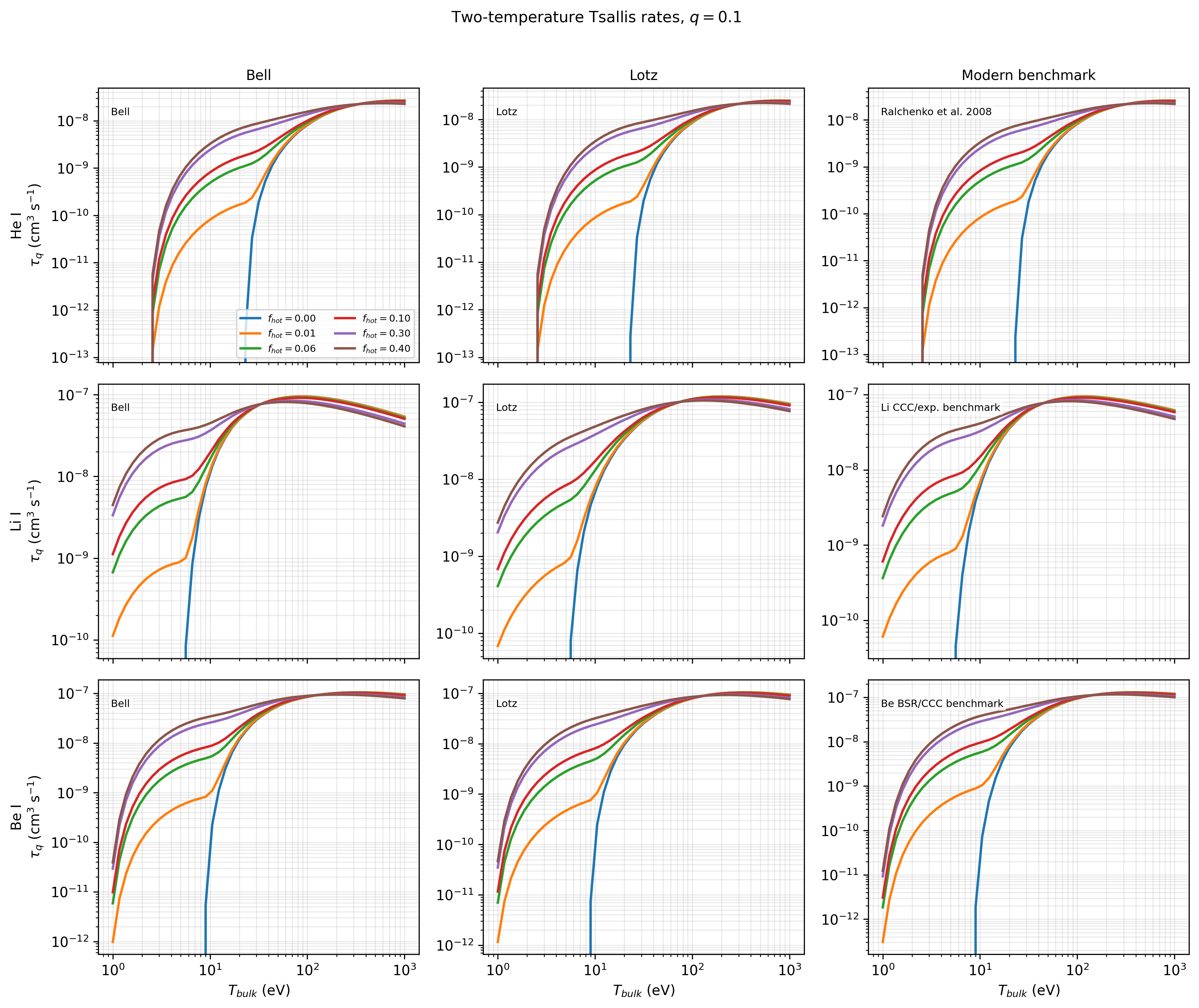}
\caption{Rate coefficients at $q=0.1$ (extreme sub-extensive
limit), arranged as in Fig.~\ref{fig:q1}: species by rows and atomic-data
input by columns. The EEDF support is now
$E\le E_{\max}\simeq 1.11\,T$, barely above bulk thermal energy.
Suppression of $\tau_q$ at $T_{\mathrm{bulk}}=1$~eV reaches
$\sim 10^4$ for He, $\sim 10^2$ for Be, and $\sim 10$ for Li.
The Bell, Lotz, and modern-benchmark columns show that the same
sub-extensive suppression persists independently of the cross-section input.}
\label{fig:q01}
\end{figure*}

At $q=0.1$ (Fig.~\ref{fig:q01}) the cut-off approaches the
threshold itself and the suppression becomes extreme. Over most
of the plotted temperature range the He rate drops by orders of
magnitude, while the Li rate, whose threshold sits inside the
retained EEDF support for all $T\gtrsim 5$~eV, is still
appreciable. Be behaves as a milder version of He. This is a
concrete illustration of the general rule that the
non-Maxwellian correction is controlled not only by $q$ itself
but also by the position of the ionization threshold relative to
the effective EEDF support.

\subsection{Super-extensive regime $q>1$}

\begin{figure*}[t]
\centering
\includegraphics[width=0.95\linewidth]{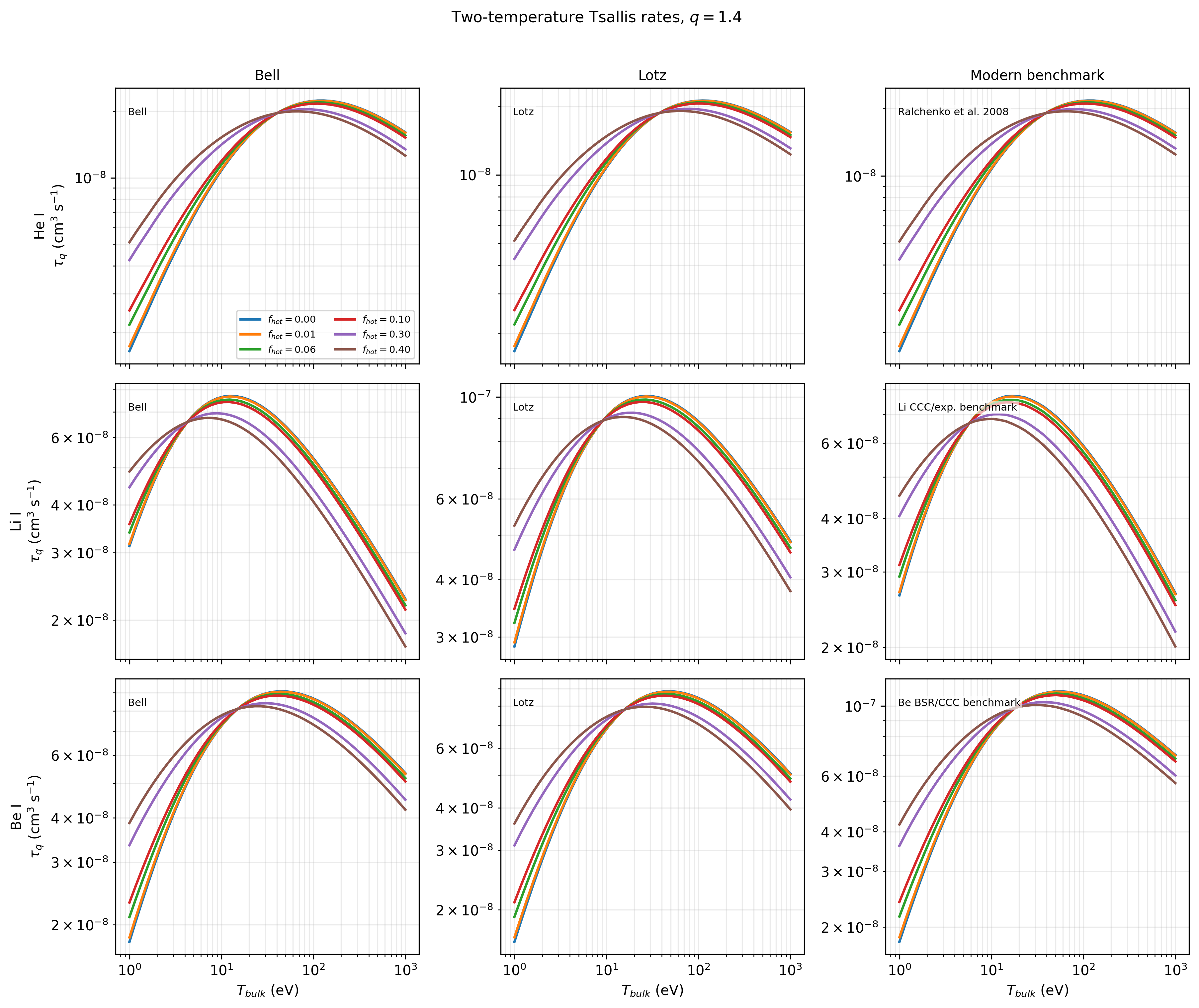}
\caption{Rate coefficients at $q=1.4$ ($\kappa=2.5$), with
species by rows and atomic-data input by columns. The Maxwellian exponential
tail is replaced by the algebraic decay
$f_q\propto[1+(E/T)/\kappa]^{-\kappa}$ of Eq.~\eqref{eq:fkappa}.
At $T_{\mathrm{bulk}}\sim 1$~eV the rate is enhanced relative to the $q=1$
baseline by one to two orders of magnitude for He, an order of magnitude for
Be, and a factor of a few for Li. The separated columns show the same EEDF
effect computed with Bell, Lotz, and the species-resolved modern benchmark.
Note: $q=1.4$ sits at the finite-mean-energy boundary $q=7/5$ and should be
interpreted as a borderline sensitivity test (Sec.~\ref{sec:tsallis-results}).}
\label{fig:q14}
\end{figure*}

For $q>1$ the trend reverses. A power-law tail develops and
populates the suprathermal sector more heavily than a
Maxwellian. Figure~\ref{fig:q14} shows the result for $q=1.4$
(equivalent to $\kappa=2.5$). At low bulk temperatures the rate
is enhanced relative to the $q=1$ Maxwellian, most strongly for
He because its higher threshold makes the rate most sensitive
to energetic electrons. At high temperatures the enhancement
narrows: the Maxwellian itself populates the relevant energies
and the $\kappa$ tail no longer dominates.

\begin{figure*}[t]
\centering
\includegraphics[width=0.95\linewidth]{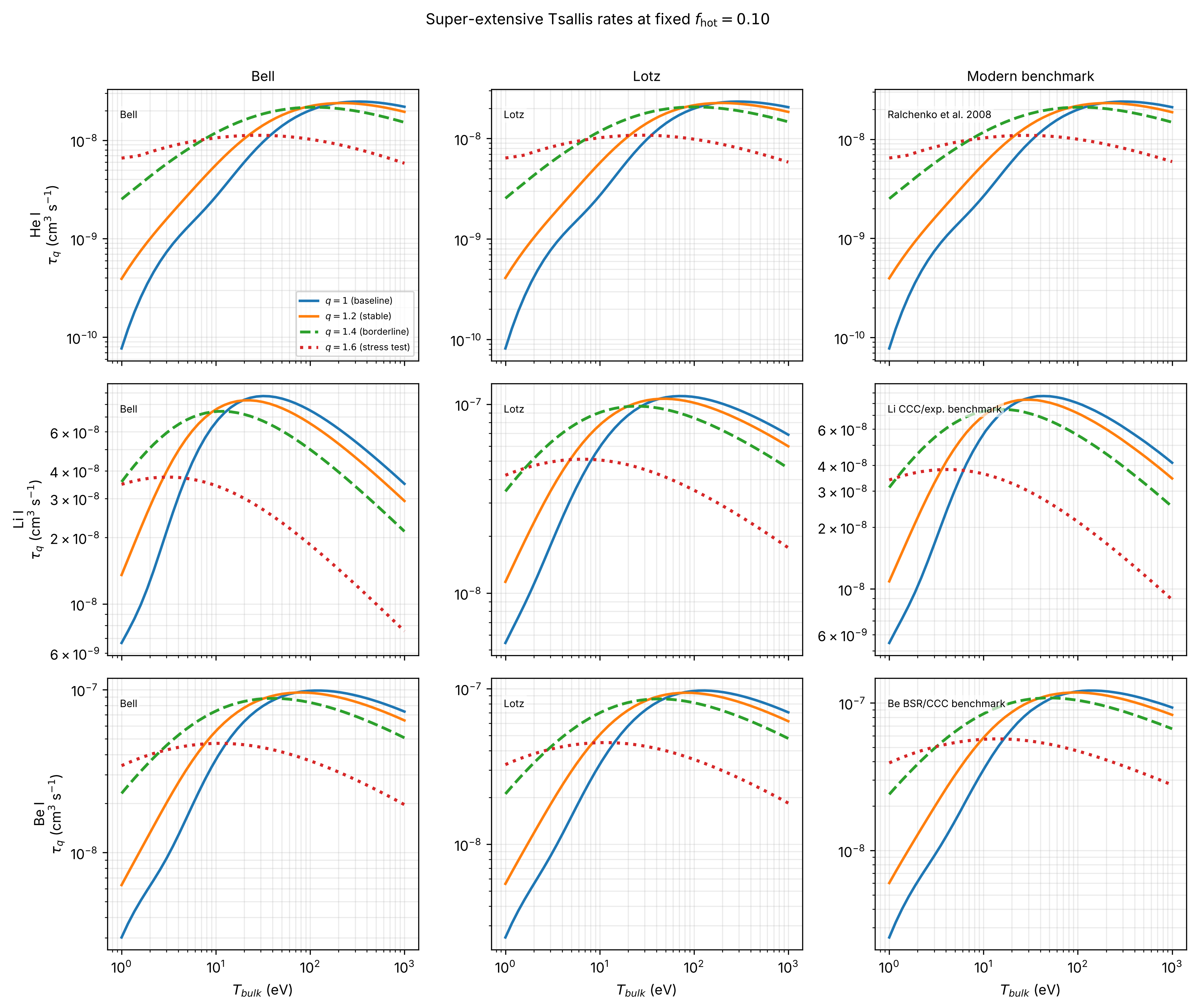}
\caption{Rate coefficients in the super-extensive regime, split by
species (rows) and atomic-data input (columns). Colours label
$q=1,1.2,1.4,1.6$, and every curve uses the same
$f_{\mathrm{hot}}=0.10$ with
$T_{\mathrm{hot}}=10T_{\mathrm{bulk}}$. Thus the comparison isolates the
effect of $q$; the $q=1$ curve is the two-temperature Maxwellian baseline,
not the strict single-Maxwellian case. \RevIII{Line style encodes the
reliability hierarchy established in Secs.~\ref{sec:cutoff}
and~\ref{sec:tsallis-results}: solid for the quantitatively stable cases
(the $q=1$ baseline and $q=1.2$), dashed for the borderline
finite-mean-energy case ($q=1.4$), and dotted for the cutoff-sensitive
heavy-tail stress test ($q=1.6$).} The modern benchmark column uses
Ralchenko \textit{et~al.} for He, the Li CCC/experimental digitization for
Li, and the Be BSR/CCC digitization for Be. The $q=1.2$ result is
quantitatively stable, $q=1.4$ is a borderline finite-mean-energy case, and
$q=1.6$ is shown only as a cutoff-sensitive heavy-tail stress test.}
\label{fig:qgt1}
\end{figure*}

Figure~\ref{fig:qgt1} compares the super-extensive family at
the same hot-electron fraction, $f_{\mathrm{hot}}=0.10$, for
every curve. Consequently the ordering isolates the effect of
$q$ rather than mixing a change in $q$ with a change in the
two-temperature composition. At low $T$, larger $q$ (smaller
$\kappa_{\mathrm{T}}$) gives a heavier tail and a larger rate.
Relative to the two-temperature $q=1$ baseline, the kinetically
well-defined $q=1.2$ case enhances the Bell He rate by factors
of about $5.1$ at $1$~eV and $2.1$ at $10$~eV; the
corresponding effects for Li and Be are smaller. The $q=1.4$
curve lies at the finite-mean-energy boundary and is a
borderline sensitivity case. The $q=1.6$ curve lies in the
normalizable but infinite-mean-energy window and, as shown in
Sec.~\ref{sec:cutoff}, is also cutoff-sensitive; it is
displayed only as an extreme stress test.

\subsection{Two complementary departures from a Maxwellian EEDF}

In the Tsallis formulation used here, the microscopic collision
physics is left unchanged: only the statistical weight $f(E)$ with
which each energy enters the rate integral changes. The
comparison with a standard Maxwellian calculation is therefore
naturally made at the level of $\langle\sigma v\rangle$, not at
the level of $\sigma(E)$. The two branches considered in the
present paper bracket two physically relevant and opposite
departures from a Maxwellian EEDF.

The super-extensive branch $q>1$ produces the $\kappa$-like
EEDF Eq.~\eqref{eq:fkappa}, with the explicitly defined tail
index $\kappa_{\mathrm{T}}=1/(q-1)$. The low-temperature
enhancement of $\tau_q$
displayed in Fig.~\ref{fig:qgt1} is then the direct kinematic
consequence of the heavier-than-Maxwellian high-energy tail: at
$T_{\mathrm{bulk}}\ll I_p/k_B$, the Maxwellian EEDF supplies
exponentially few electrons above threshold, whereas the
algebraic Tsallis tail still has support there. The
enhancement scales with the relative weight of the suprathermal
tail and with $I_p/k_BT$, as the species-dependent ordering of
Fig.~\ref{fig:qgt1} confirms. This regime is the one of practical
relevance for non-thermal coronal and flare
plasmas~\cite{Pierrard2010,Dudik2017,Lorincik2020,Jeffrey2024}
and for fusion-edge plasmas with suprathermal
populations~\cite{Hartfuss1997,HansenShlyaptseva2004,Brix2010}.

The sub-extensive branch $q<1$ has no direct $\kappa$ counterpart
because the EEDF support is finite. In the present calculations
this branch is used as a phenomenological model of tail
depletion: situations where suprathermal electrons are removed
faster than they are generated by Coulomb relaxation---e.g.\ by
fast inelastic losses, by cold-wall boundary conditions, or by
strong inelastic-collisional sinks above a characteristic
energy. The associated suppression of $\tau_q$, displayed in
Figs.~\ref{fig:q07}--\ref{fig:q01}, scales again with $I_p/k_BT$
and is most severe for He.

{\color{blue}%
\subsection{Physical implications and the expected
ionization-balance response}
\label{sec:implications}

The calculated rates imply a testable expectation for a
complete ionization--recombination model, but they do not by
themselves determine an ionization balance. For helium, the
moderate $q=1.2$ tail at fixed $f_{\mathrm{hot}}=0.10$ raises
the ionization coefficient relative to the corresponding
two-temperature $q=1$ baseline by a factor of about five at
$T_{\mathrm{bulk}}=1$~eV and about two at $10$~eV. Because
ionization samples the suprathermal part of the EEDF more
strongly than many recombination channels, this enhancement is
expected to move the neutral-He abundance toward lower bulk
temperatures. The magnitude and detailed location of that
shift, however, require species-consistent recombination
coefficients and a full collisional-radiative charge-state
network. The present work supplies the ionization input and a
scaling hypothesis; it does not claim to have calculated the
charge-state fractions themselves.
}

\section{Conclusions}
\label{sec:conclusion}

We have computed non-Maxwellian electron-impact
single-ionization rate coefficients for neutral He, Li, and Be
by inserting the recommended Bell cross
sections~\cite{Bell1983} into a two-temperature Tsallis EEDF
with properly per-branch normalization. The calculation has
been positioned with respect to two complementary references:
at the level of the microscopic cross section, the classical
Lotz formula~\cite{Lotz1967,Lotz1968}; and at the level of
the rate coefficient itself, the standard Maxwellian
compilations~\cite{Kato1991,ArnaudRothenflug1985} in the
strict limit $(q,f_{\mathrm{hot}})=(1,0)$. Four conclusions
emerge:
\begin{enumerate}
  \item The Bell analytic coefficients must be implemented with
        the negative signs of $B_1,B_2$ for He\,\textsc{i} and of
        $B_1$ for Li\,\textsc{i} (Table~\ref{tab:params}) to
        reproduce Bell's recommended He\,\textsc{i} plate.
  \item The cross-section model dependence is
        strongly species-dependent: Bell, Lotz and Ralchenko agree closely for He\,\textsc{i}, while Bell and Lotz agree within
        $\lesssim 7\%$ on $\tau_M$ for He, within $\lesssim 17\%$
        for Be, and differ by up to $+95\%$ at $T=1$~keV for Li.
        The ordering is transmitted from $\sigma(E)$ to
        $\tau_q(T)$ without qualitative change.
  \item The non-Maxwellian correction scales with $I_p/k_B T$.
        Sub-extensive distributions ($q<1$) suppress
        ionization---strongly for He, moderately for Be, mildly
        for Li. Super-extensive distributions ($q>1$) enhance
        ionization at low $T$, with the same species ordering.
  \item The super-extensive branch is $\kappa$-like, with the
        tail index $\kappa_{\mathrm{T}}=1/(q-1)$ defined
        specifically for Eq.~\eqref{eq:fkappa}. It is not
        identical to the standard plasma form by a simple
        parameter shift. The values
        $q\in\{1.2,\,1.4,\,1.6\}$ span progressively heavier
        algebraic tails rather than constituting a direct
        calibration to a unique observational $\kappa$
        convention. The sub-extensive branch $q<1$ has no
        $\kappa$ counterpart and is used here as a
        phenomenological model of tail depletion.
  \item {\color{red}Modern benchmark cross sections are propagated
        numerically for all three species: Ralchenko \textit{et~al.} for
        He\,\textsc{i}, a Bray-type CCC/experimental benchmark for
        Li\,\textsc{i}, and a BSR/CCC benchmark for Be\,\textsc{i}. The
        benchmark-Bell differences in the strict Maxwellian rate are small
        for He, moderate for Li, and larger for Be above the peak. This
        comparison identifies which species are most sensitive to the
        microscopic atomic input.}
        {\color{blue}The physical consequence is concrete
        (Sec.~\ref{sec:implications}): a kinetically well-defined
        $q=1.2$ tail raises the He ionization rate by a factor of
        about five at $1$~eV relative to the equal-mixture $q=1$
        baseline and is expected, in a complete
        ionization--recombination model, to move neutral He toward
        lower bulk temperature,
        making the shape of the EEDF---within the benchmarked atomic-data
        envelope---the dominant modelling uncertainty for light neutrals in
        non-thermal plasmas.}
\end{enumerate}

\paragraph*{Limitations.}
The present calculation is restricted to single ionization of
neutral light atoms; excitation-autoionization
channels~\cite{Bryans2009} and the full
collisional-radiative coupling that determines line
intensities in a real plasma~\cite{Hahn2017,Dere2007} are
beyond its scope. The Tsallis entropic index $q$ and the
hot-electron fraction $f_{\mathrm{hot}}$ are treated as free
phenomenological parameters rather than as quantities derived
from a kinetic equation; in any specific astrophysical or
laboratory setting they would have to be inferred from
observations or from a self-consistent simulation. The Lotz
cross section used as a comparator is the one-shell form
which omits inner-subshell contributions that may matter at
high incident energy. Finally, as detailed in
Sec.~\ref{sec:cutoff}, the rate-coefficient values reported
for $q=1.6$ at the lowest plotted temperatures are subject to
significant cutoff sensitivity (up to about $25\%$ in the
reported test) and should be
read as qualitative tendencies rather than as quantitatively
reliable numbers; they nevertheless preserve the species
ordering and the sign of the non-Maxwellian effect. Within
these limits the present results provide a controlled
sensitivity test of EII rate coefficients of light neutrals
against both cross-section model uncertainty and
non-Maxwellian EEDF deformation.

{\color{red}Future work may extend the present analysis by incorporating
inner-shell and excitation-autoionization channels and by embedding the
calculated rate coefficients in a complete collisional-radiative model.
Such an extension would permit a direct quantitative assessment of
non-Maxwellian effects on spectral-line intensities and charge-state
distributions.}
{\color{RoundTwo}It is useful to state explicitly which physical
ingredients such a model would have to supply beyond the ionization
coefficients computed here, since this is precisely what delimits the
applicability of the present results. A closed ionization--recombination
description of the same light neutrals requires, in addition:
(i)~radiative recombination rate coefficients, which must themselves be
re-averaged over the non-Maxwellian EEDF rather than taken from
Maxwellian fits, since the recombination integrand weights the
low-energy part of $f(E)$ that the Tsallis deformation also modifies;
(ii)~dielectronic recombination, weak for these low-charge ions but not
strictly negligible at the upper end of the temperature range considered;
(iii)~three-body recombination, which becomes competitive at the high
electron densities typical of laboratory discharges and divertor plasmas;
(iv)~charge-exchange reactions with neutral hydrogen, which can dominate
the recombination balance under fusion-edge conditions;
(v)~excitation-autoionization and inner-shell ionization channels, omitted
from the present single-ionization treatment; and (vi)~the coupling of all
charge states---including metastable populations---through a
time-dependent collisional-radiative network, because the quasi-static
coronal approximation may fail when the EEDF departs strongly from a
Maxwellian. The present work supplies only the first of these ingredients;
the shift in ionization balance discussed in Sec.~\ref{sec:implications}
is therefore a scaling expectation, and becomes a quantitative prediction
only once the remaining terms are evaluated on the same EEDF.}

\begin{acknowledgments}
{\color{red}The author thanks the anonymous referees for their
careful reading and constructive comments.}
\end{acknowledgments}

\appendix

\section{Derivation of the normalization $A_q(T)$}
\label{app:Aq}

This appendix derives the closed-form expression
Eq.~\eqref{eq:Aq} for the normalization constant of the
Tsallis EEDF Eq.~\eqref{eq:fq}. The derivation uses only the
Beta-function identity
\begin{equation}
  B(\alpha,\beta) \;=\;
  \int_{0}^{1} t^{\alpha-1}(1-t)^{\beta-1}\,dt
  \;=\; \frac{\Gamma(\alpha)\,\Gamma(\beta)}
              {\Gamma(\alpha+\beta)},
  \label{eq:Beta}
\end{equation}
valid for $\Re\alpha>0$, $\Re\beta>0$, and the closely related
identity
\begin{equation}
  \int_{0}^{\infty} \frac{u^{\alpha-1}}{(1+u)^{\alpha+\beta}}\,du
  \;=\; B(\alpha,\beta),
  \label{eq:BetaB}
\end{equation}
obtained from Eq.~\eqref{eq:Beta} by the substitution
$t=u/(1+u)$.

\subsection{Sub-extensive branch ($q<1$)}

For $q<1$ the EEDF Eq.~\eqref{eq:fq} reads
\begin{equation}
  f_q(E;T) \;=\; A_q(T)\,\sqrt{E}\,
  \left[1-(1-q)\,\frac{E}{T}\right]^{1/(1-q)},
  \label{app:fq-sub}
\end{equation}
with compact support $0\le E\le E_{\max}=T/(1-q)$. The
normalization condition is
\begin{equation}
  1 \;=\; \int_{0}^{E_{\max}} f_q(E;T)\,dE
  \;=\; A_q(T)\,\mathcal{I}_{<}(T),
  \label{app:norm-sub}
\end{equation}
with
\begin{equation}
  \mathcal{I}_{<}(T) \;=\;
  \int_{0}^{E_{\max}} \sqrt{E}\,
  \left[1-(1-q)\,\frac{E}{T}\right]^{1/(1-q)} dE.
  \label{app:I-sub}
\end{equation}
Introduce the dimensionless variable
\begin{equation}
  t \;=\; (1-q)\,\frac{E}{T},
  \quad
  E \;=\; \frac{T}{1-q}\,t,
  \quad
  dE \;=\; \frac{T}{1-q}\,dt.
\end{equation}
The bounds of integration become $t\in[0,1]$ and the integrand
becomes
\begin{align}
  \sqrt{E}\,[1-(1-q)E/T]^{1/(1-q)}
  &\;=\; \left(\frac{T}{1-q}\right)^{\!1/2}\!t^{1/2}
         (1-t)^{1/(1-q)}.
\end{align}
The integral Eq.~\eqref{app:I-sub} therefore becomes
\begin{align}
  \mathcal{I}_{<}(T)
  &\;=\; \left(\frac{T}{1-q}\right)^{\!3/2}
         \int_{0}^{1} t^{1/2}(1-t)^{1/(1-q)} dt \nonumber\\
  &\;=\; \left(\frac{T}{1-q}\right)^{\!3/2}
         B\!\left(\tfrac{3}{2},\,\tfrac{1}{1-q}+1\right),
  \label{app:I-sub-Beta}
\end{align}
using Eq.~\eqref{eq:Beta} with $\alpha=3/2$ and
$\beta=1/(1-q)+1$. Both arguments are positive for $0<q<1$,
which is the entire admissible range, so the integral converges
without restriction.

Using $\Gamma(3/2)=\tfrac{1}{2}\sqrt{\pi}$, the Beta function
expands to
\begin{equation}
  B\!\left(\tfrac{3}{2},\tfrac{1}{1-q}+1\right)
  \;=\; \frac{\tfrac{1}{2}\sqrt{\pi}\;
              \Gamma\!\left(\tfrac{1}{1-q}+1\right)}
             {\Gamma\!\left(\tfrac{1}{1-q}+\tfrac{5}{2}\right)}.
\end{equation}
Substituting in Eq.~\eqref{app:norm-sub} and solving for
$A_q(T)$,
\begin{equation}
  A_q(T) \;=\;
  \frac{2}{\sqrt{\pi}}\,T^{-3/2}\,(1-q)^{3/2}\,
  \frac{\Gamma\!\left(\tfrac{1}{1-q}+\tfrac{5}{2}\right)}
       {\Gamma\!\left(\tfrac{1}{1-q}+1\right)},
  \quad q<1,
  \label{app:Aq-sub}
\end{equation}
which is the upper line of Eq.~\eqref{eq:Aq}.

\subsection{Super-extensive branch ($q>1$)}

For $q>1$ the EEDF Eq.~\eqref{eq:fq} reads
\begin{equation}
  f_q(E;T) \;=\; A_q(T)\,\sqrt{E}\,
  \left[1+(q-1)\,\frac{E}{T}\right]^{-1/(q-1)},
  \label{app:fq-sup}
\end{equation}
with support $E\in[0,\infty)$. The normalization condition is
\begin{equation}
  1 \;=\; A_q(T)\,\mathcal{I}_{>}(T),
\end{equation}
with
\begin{equation}
  \mathcal{I}_{>}(T) \;=\;
  \int_{0}^{\infty} \sqrt{E}\,
  \left[1+(q-1)\,\frac{E}{T}\right]^{-1/(q-1)} dE.
  \label{app:I-sup}
\end{equation}
Introduce
\begin{equation}
  u \;=\; (q-1)\,\frac{E}{T},
  \quad
  E \;=\; \frac{T}{q-1}\,u,
  \quad
  dE \;=\; \frac{T}{q-1}\,du.
\end{equation}
The bounds of integration become $u\in[0,\infty)$ and the
integrand becomes
\begin{equation}
  \sqrt{E}\,[1+(q-1)E/T]^{-1/(q-1)}
  \;=\; \left(\frac{T}{q-1}\right)^{\!1/2}\!u^{1/2}
        (1+u)^{-1/(q-1)}.
\end{equation}
Therefore
\begin{equation}
  \mathcal{I}_{>}(T)
  \;=\; \left(\frac{T}{q-1}\right)^{\!3/2}
        \int_{0}^{\infty}
        \frac{u^{1/2}}{(1+u)^{1/(q-1)}}\,du.
  \label{app:I-sup-step}
\end{equation}
Identifying the integrand with Eq.~\eqref{eq:BetaB} with
$\alpha-1=1/2$ and $\alpha+\beta=1/(q-1)$, i.e.\ $\alpha=3/2$
and $\beta=1/(q-1)-3/2$, gives
\begin{equation}
  \int_{0}^{\infty}\!\!\frac{u^{1/2}}{(1+u)^{1/(q-1)}}\,du
  \;=\; B\!\left(\tfrac{3}{2},\,\tfrac{1}{q-1}-\tfrac{3}{2}\right),
\end{equation}
which converges if and only if $\beta>0$, i.e.\
$1/(q-1)>3/2$, i.e.\
\begin{equation}
  q \;<\; \frac{5}{3}
  \quad\Longleftrightarrow\quad
  \kappa \;>\; \frac{3}{2}.
  \label{app:q-bound}
\end{equation}
This is the \emph{normalization} bound on the admissible
Tsallis parameter, and the equivalent bound $\kappa>3/2$
familiar from the $\kappa$-distribution
literature~\cite{Livadiotis2017book}. For $q\ge 5/3$ the EEDF
itself fails to integrate to a finite value and is not usable.

A second, stricter bound governs the existence of a finite
mean kinetic energy. Multiplying the integrand of
Eq.~\eqref{app:I-sup} by an extra factor of $E$ (or
equivalently of $u$) and repeating the analysis identifies the
moment integral with $B(5/2,\,1/(q-1)-5/2)$, which converges
only when $1/(q-1)>5/2$, i.e.\
\begin{equation}
  q \;<\; \frac{7}{5}
  \quad\Longleftrightarrow\quad
  \kappa_{\mathrm{T}} \;>\; \frac{5}{2}.
  \label{app:q-energy-bound}
\end{equation}
For $q\in[7/5,\,5/3)$ the EEDF is normalizable but does not
admit a finite mean energy: in this regime $f_q$ should be
read as a phenomenological heavy-tail model rather than as the
equilibrium distribution of a kinetic temperature. For the
Bell and Lotz forms, $\sigma(E)\sim\ln(E)/E$ at high energy.
Since $v(E)\propto E^{1/2}$ and
$f_q(E)\propto E^{1/2-\kappa_{\mathrm{T}}}$, the complete rate
integrand behaves as $E^{-\kappa_{\mathrm{T}}}\ln E$. The rate
therefore converges for $\kappa_{\mathrm{T}}>1$ (equivalently
$q<2$). The stronger EEDF-normalization requirement
$\kappa_{\mathrm{T}}>3/2$ already guarantees this condition, so
the Bell and Lotz rate coefficients are finite throughout the
normalizable range $q<5/3$. The two distribution-moment bounds
Eqs.~\eqref{app:q-bound} and
\eqref{app:q-energy-bound} are summarised in
Table~\ref{tab:qbounds}.

\begin{table}[h]
\centering
\caption{Bounds on the Tsallis entropic index $q$ in the
super-extensive branch. The first bound is required for the
EEDF to be normalizable; the second, stricter bound is
required for the EEDF to have a finite mean kinetic energy.
For Bell and Lotz, the rate-coefficient integral is finite
throughout the EEDF-normalizable range $q<5/3$.}
\label{tab:qbounds}
\begin{ruledtabular}
\begin{tabular}{l c c}
quantity & $q$ bound & $\kappa_{\mathrm{T}}$ bound \\ \hline
EEDF normalization                & $q<5/3$ & $\kappa_{\mathrm{T}}>3/2$ \\
finite mean kinetic energy        & $q<7/5$ & $\kappa_{\mathrm{T}}>5/2$ \\
$\tau_q(T)$ for Bell or Lotz $\sigma$ & $q<5/3$ & $\kappa_{\mathrm{T}}>3/2$ \\
\end{tabular}
\end{ruledtabular}
\end{table}

Expanding the Beta function with $\Gamma(3/2)=\sqrt{\pi}/2$,
\begin{equation}
  B\!\left(\tfrac{3}{2},\tfrac{1}{q-1}-\tfrac{3}{2}\right)
  \;=\; \frac{\tfrac{1}{2}\sqrt{\pi}\;
              \Gamma\!\left(\tfrac{1}{q-1}-\tfrac{3}{2}\right)}
             {\Gamma\!\left(\tfrac{1}{q-1}\right)},
\end{equation}
and substituting,
\begin{equation}
  A_q(T) \;=\;
  \frac{2}{\sqrt{\pi}}\,T^{-3/2}\,(q-1)^{3/2}\,
  \frac{\Gamma\!\left(\tfrac{1}{q-1}\right)}
       {\Gamma\!\left(\tfrac{1}{q-1}-\tfrac{3}{2}\right)},
  \qquad 1<q<5/3.
  \label{app:Aq-sup}
\end{equation}
which is the lower line of Eq.~\eqref{eq:Aq}.

\subsection{Maxwellian limit and consistency check}

In the Maxwellian limit $q\to 1^{\pm}$ both
Eq.~\eqref{app:Aq-sub} and Eq.~\eqref{app:Aq-sup} should reduce
to the standard Maxwell--Boltzmann prefactor
$A_1(T)=\tfrac{2}{\sqrt{\pi}}T^{-3/2}$ of Eq.~\eqref{eq:fM}.
This follows from the Stirling-type asymptotic
identity~\cite{AbramowitzStegun}
\begin{equation}
  \lim_{n\to\infty}\, n^{-c}\,\frac{\Gamma(n+c)}{\Gamma(n)}
  \;=\; 1,
  \label{app:stirling}
\end{equation}
which holds for any constant $c\in\mathbb{R}$.

Consider first the sub-extensive branch
Eq.~\eqref{app:Aq-sub}. Set
\begin{equation}
  n \;=\; \frac{1}{1-q}+1,
  \qquad c \;=\; \tfrac{3}{2}.
\end{equation}
Then $n+c=1/(1-q)+5/2$, so
$\Gamma(n+c)/\Gamma(n)\sim n^{3/2}$ as $q\to 1^{-}$
(equivalently $n\to\infty$), by Eq.~\eqref{app:stirling}.
Substituting into Eq.~\eqref{app:Aq-sub} gives
\begin{equation}
  A_q(T)
  \;\sim\;
  \frac{2}{\sqrt{\pi}}\,T^{-3/2}\,(1-q)^{3/2}\,n^{3/2}
  \quad (q\to 1^{-}).
\end{equation}
Because $n=(2-q)/(1-q)$, the product $(1-q)^{3/2}\,n^{3/2}$
equals $(2-q)^{3/2}$, which approaches unity as $q\to 1^{-}$.
Hence
\begin{equation}
  \lim_{q\to 1^{-}} A_q(T)
  \;=\; \frac{2}{\sqrt{\pi}}\,T^{-3/2},
\end{equation}
which is precisely the Maxwellian prefactor of
Eq.~\eqref{eq:fM}.

The same argument applies to the super-extensive branch
Eq.~\eqref{app:Aq-sup}. Setting now $n=1/(q-1)-3/2$ and
$c=3/2$ in Eq.~\eqref{app:stirling}, one finds
$\Gamma(n+c)/\Gamma(n)\sim n^{3/2}$ as $q\to 1^{+}$, and the
combination $(q-1)^{3/2}\,n^{3/2}$ again tends to unity. The
Maxwellian limit is therefore approached from above as well.
The two branches join continuously at $q=1$, and
Eq.~\eqref{eq:Aq} can be used as a single closed-form
prefactor across the entire admissible range $0<q<5/3$.

\subsection{Relation to common plasma-$\kappa$ forms}

The tail index $\kappa_{\mathrm{T}}=1/(q-1)$ used here is the
absolute value of the exponent in Eq.~\eqref{app:fq-sup}. A
common plasma convention instead writes
\begin{equation}
  f_{\kappa_{\mathrm{p}}}(E) \;\propto\; \sqrt{E}\,
  \left[1+\frac{E}{(\kappa_{\mathrm{p}}-3/2)\,T_\kappa}
  \right]^{-\kappa_{\mathrm{p}}-1}.
  \label{app:kappa-plasma}
\end{equation}
Although the exponents can be related formally by
$\kappa_{\mathrm{T}}=\kappa_{\mathrm{p}}+1$, the scale inside
the bracket also changes. The two distributions are therefore
not identical under a trivial symbol shift unless the
temperature parameter is transformed at the same time. All
$\kappa$ values quoted in this paper are explicitly
$\kappa_{\mathrm{T}}$ values for Eq.~\eqref{eq:fkappa}.

\end{document}